\documentclass[12pt]{article}
\usepackage{psfrag,graphicx}
\def\baselinestretch{1.4}
\usepackage{array}
\catcode`@=12
\usepackage[square,comma,sort&compress]{natbib}
\usepackage{graphicx}
\usepackage[figuresright]{rotating}
\usepackage{array}
\newcommand{\beq}{\begin{equation}}
\newcommand{\eeq}{\end{equation}}
\newcommand{\bea}{\begin{eqnarray}}
\newcommand{\eea}{\end{eqnarray}}

\DeclareMathAlphabet{\mathsc}{OT1}{cmr}{m}{sc}
\newcommand{\TM} {\texttt{TM}}
\newcommand{\TMs} {\texttt{TMs }}
\newcommand{\Sbl}  {\mathsc{sbl}}

\newcommand{\Lsnd} {\mathsc{lsnd}}

\newcommand{\gof}{g.o.f.}
\newcommand{\dml}{\Delta m^2_\Lsnd}

\newcommand{\CL}   {C.L.}
\newcommand{\dof}  {d.o.f.}

\newcommand{\eVq}  {\rm{eV}^2}
\newcommand{\Sol}  {\mathsc{sol}}
\newcommand{\Atm}  {\mathsc{atm}}

\newcommand{\Dcq}  {\Delta\chi^2}
\newcommand{\Dms}  {\Delta m^2_\Sol}
\newcommand{\Dma}  {\Delta m^2_\Atm}

\newcommand{\AHEP}{Instituto de F\'{\i}sica Corpuscular --
  C.S.I.C./Universitat de Val{\`e}ncia \\
  Edificio Institutos de Paterna, Apt 22085,
  E--46071 Val{\`e}ncia, Spain\\
 http://ific.uv.es/\~{}ahep}
\newcommand{\hawaii}{
Department of Physics and Astronomy, University of Hawaii, Manoa\\
2505 Correa Road, Honolulu, Hawaii 96822 USA\\}
\newcommand{\snocc}{SNO$_\mathrm{CC}^\mathrm{rate}$ }
\newcommand{\snotot}{SNO$_\mathrm{CC,NC}^\mathrm{SP,DN}$ }
\def\lsim{\raise0.3ex\hbox{$\;<$\kern-0.75em\raise-1.1ex\hbox{$\sim\;$}}}
\def\gsim{\raise0.3ex\hbox{$\;>$\kern-0.75em\raise-1.1ex\hbox{$\sim\;$}}}
\def\e6{$E(6)$}
\def\10{$SO(10)$}
\def\21{$SU(2) \otimes U(1) $}

\def\422{$SU(4) \otimes SU(2) \otimes SU(2)$}
\def\321{$SU(3) \otimes SU(2) \otimes U(1)$}
\def\ne{\hbox{$\nu_e$ }}
\def\nm{\hbox{$\nu_\mu$ }}
\def\bne{\hbox{$\bar\nu_e$ }}
\def\bnm{\hbox{$\bar\nu_\mu$ }}
\def\bnt{\hbox{$\bar\nu_\tau$ }}
\def\nt{\hbox{$\nu_\tau$ }}

\def \nbb {$\beta\beta_{0\nu}$ }

\let\vev\VEV

\newcommand{\AmS}{{\protect\the\textfont2
 A\kern-.1667em\lower.5ex\hbox{M}\kern-.125emS}}

\begin{document}

\clearpage
\thispagestyle{empty}
\begin{center}

\textbf{\large Neutrino Properties Before and After KamLAND}

S. Pakvasa~$^a$ and J.~W.~F. Valle~$^b$\\
$^a$ \hawaii 
$^b$ \AHEP

\end{center}

\begin{abstract}
  
  We review neutrino oscillation physics, including the determination
  of mass splittings and mixings from current solar, atmospheric,
  reactor and accelerator neutrino data.  A brief discussion is given
  of cosmological and astrophysical implications. Non-oscillation
  phenomena such as neutrinoless double beta decay would, if
  discovered, probe the absolute scale of neutrino mass and also
  reveal their Majorana nature.
  Non-oscillation descriptions in terms of spin-flavor precession
  (SFP) and non-standard neutrino interactions (NSI) currently provide
  an excellent fit of the solar data.
  However they are at odds with the first results from the KamLAND
  experiment which imply that, despite their theoretical interest,
  non-standard mechanisms can only play a sub-leading role in the
  solar neutrino anomaly.
  Accepting the LMA-MSW solution, one can use the current solar
  neutrino data to place important restrictions on non-standard
  neutrino properties, such as neutrino magnetic moments.
  Both solar and atmospheric neutrino data can also be used to place
  constraints on neutrino instability as well as the more exotic
  possibility of $CPT$ and Lorentz Violation.
  We illustrate the potential of future data from experiments such as
  KamLAND, Borexino and the upcoming neutrino factories in
  constraining non-standard neutrino properties.

\vfill
\end{abstract}
 \newpage
 %{\footnotesize \renewcommand \baselinestretch{1} \tableofcontents}
 \tableofcontents
 \newpage

\section{Introduction}
\label{sec:introduction}

Since the early Davis experiment using the geochemical method to
detect solar neutrinos via the \ne + ${}^{37}$Cl $\to {}^{37}$Ar +
$e^-$ reaction at Homestake~\cite{Cleveland:nv}, solar neutrino
research has gone a long way to become now a mature field.  The
subsequent Gallex~\cite{Hampel:1998xg},
Sage~\cite{Abdurashitov:1999bv} and GNO~\cite{Altmann:2000ft}
experiments have not only confirmed the consistency of the basic
elements of solar energy generation, but also established that the
deficit seen in the chlorine experiment also exists in the reaction
\ne + ${}^{71}$Ga $\to ^{71}$Ge + $e^-$~\cite{Altmann:2000ft}.  Direct
detection with Cerenkov techniques using $\nu_e e$ scattering on water
at Super-K \cite{Fukuda:2002pe}, and heavy water at SNO
\cite{Ahmad:2002jz,Ahmad:2001an} has given a robust confirmation that
the number of solar neutrinos detected in these underground
experiments is less than expected from theories of energy generation
in the sun \cite{Bahcall}.  Especially relevant is the sensitivity of
the SNO experiment to the neutral current (NC).
Altogether these experiments provide a solid evidence for solar
neutrino conversions and, therefore, for physics beyond the Standard
Model (SM).  Current data indicate that the mixing angle is
large~\cite{Maltoni:2002ni}, the best description being given by the
LMA-MSW solution~\cite{Wolfenstein:1977ue}, already hinted previously
from the flat Super-K recoil electron spectra~\cite{firstLMA}. We will
briefly describe the results of the analysis of solar neutrino data
and the resulting parameters in Sec.~\ref{sec:solar-neutr}.

%%%%%%% atm %%%%%

Atmospheric neutrinos are produced in hadronic showers initiated by
cosmic-ray collisions with air in the upper atmosphere~\footnote{See
  Ref.~\cite{Maltoni:2002ni} for an extensive list of experimental
  references}. They have been observed in several experiments
\cite{atm}.
Although individual $\nu_\mu$ or $\nu_e$ fluxes are only known to
within $20-30\%$ accuracy, their ratio is predicted to within $5\%$
over energies varying from 0.1~GeV to tens of~GeV~\cite{atmfluxes}.
The long-standing discrepancy between the predicted and measured
$\mu/e$ ratio of the muon-type ($\nu_\mu + \bar{\nu}_\mu$) over the
e-type ($\nu_e+\bar{\nu}_e$) atmospheric neutrino fluxes, has shown up
both in water Cerenkov experiments (Kamiokande, Super-K and IMB) as
well as in the iron calorimeter Soudan2 experiment. In addition, a
strong zenith-angle dependence has been found both in the sub-GeV and
multi-GeV energy range, but only for $\mu$--like events, the
zenith-angle distributions for the $e$--like being consistent with
expectation. Such zenith-angle distributions have also been recorded
for upward-going muon events in Super-K and MACRO, which are also
consistent with the \nm oscillation hypothesis. The atmospheric
neutrino data analysis is summarized in Sec.~\ref{sec:atmosph-neutr}.

%%%%%%%%%%%%%%%%%%%%%%%%%%%%%%%%%%%%%%%%%%%%%%%%%%%%%%%%%%%%%%%%%%%%%%

On the other hand, one has information on neutrino oscillations from
reactor and accelerator data, discussed in
Sec.~\ref{sec:react-accel-neutr}.
Except for the LSND experiment~\cite{LSND}, which claims evidence for
$\bar{\nu}_e$ appearance in a $\bar{\nu}_\mu$ beam, all of these
report no evidence for oscillations.  These experiments include the
short baseline disappearance experiments Bugey~\cite{bugey} and
CDHS~\cite{CDHS}, as well as the KARMEN neutrino
experiment~\cite{KARMEN}.

Particularly relevant is the non-observation of oscillations at Chooz
and Palo Verde reactors~\cite{CHOOZ}, which provides an important
restriction on the parameters $\Delta{m}^2_{32}$ and
$\sin^2(2\theta_{13})$.

Turning now to the new generation of long baseline neutrino
oscillation searches, in a recent paper~\cite{:2002dm} KamLAND has
found for the first time strong evidence for the disappearance of
neutrinos travelling from a power reactor to a far detector, located
at the Kamiokande site.
Most of the $\bar{\nu}_e$ flux incident at KamLAND comes from plants
located between $80-350$ km from the detector, making the average
baseline of about 180 kilometers, long enough to provide a sensitive
probe of the LMA-MSW solution of the solar neutrino
problem~\cite{firstLMA}. Therefore these results of the KamLAND
collaboration constitute the first test of the solar neutrino
oscillation hypothesis with terrestrial experiments and man-produced
neutrinos.
KamLAND also finds the parameters describing this disappearance in
terms of the oscillations to be consistent with what is required to
account for the solar neutrino problem.
As we will comment in Sec.~\ref{sec:impact-kamland-other} this implies
that non-standard solutions can not be leading explanation to the
solar neutrino anomaly.

On the other hand the K2K experiment has recently observed positive
indications of neutrino oscillation in a 250 km long-baseline
setup~\cite{Ahn:2002up}. The collaboration observes a reduction of
$\nu_\mu$ flux together with a distortion of the energy spectrum.  The
probability that the observed flux at Super-K is a statistical
fluctuation without neutrino oscillation is less than 1\%.

\section{Basic Neutrino Parameters}
\label{sec:basic-neutr-param}

\subsection{Neutrino Oscillation Parameters}
\label{sec:neutr-oscill-param}

Current neutrino data require three light neutrinos participating in
the oscillations. Correspondingly, the simplest structure of the
neutrino sector involves the following parameters:
\begin{itemize}
\item the solar angle $\theta_\Sol \equiv \theta_{12}$ (large, but
  substantially non-maximal) and the solar splitting $\Delta{m}^2_{21}
  \equiv \Dms$
\item the atmospheric angle $\theta_\Atm \equiv \theta_{23}$ (nearly
  maximal) and the atmospheric splitting $\Delta{m}^2_{32} \equiv \Dma
  \gg \Dms$
\item the reactor angle $\theta_{13}$ (small) 
\end{itemize}
Since in the Standard Model neutrinos are massless, their masses must
arise from some new physics. An attractive possibility is the seesaw
mechanism~\cite{seesaw79,seesaw80,seesawmajoron}.  However nothing is
presently known about whether this is the mechanism producing
neutrinos masses and, if so, what is the magnitude of the
corresponding mass scale. In fact a more general view is that neutrino
masses come from some unknown dimension-five operator
\cite{Weinberg:uk}.

In contrast, neutrino masses could well be generated at the weak
scale. One possibility is to have them induced by radiative
corrections~\cite{Zee:1980ai}.  Alternatively, neutrino masses may
have a supersymmetric origin, resulting from the spontaneous violation
of R parity~\cite{Masiero:1990uj}. In this case one is left with a
hybrid scheme where only the atmospheric scale comes from a
(weak-scale) seesaw, while the solar scale is calculable from
radiative corrections~\cite{Hirsch:2000ef}~\footnote{The idea that
  neutrino masses arise from broken R parity supersymmetry can be
  tested at collider experiments~\cite{Hirsch:2002ys}}.

Out of the three neutrino masses, only two splittings are fixed by
oscillation data.  As will be seen in Secs.~\ref{sec:solar-neutr} and
\ref{sec:atmosph-neutr}, the neutrino of mass splittings needed to fit
the observed solar and atmospheric neutrino anomalies are somewhat
hierarchical.  Depending on the sign of $\Delta m_{32}^2$ there are
three types of neutrino mass spectra which fit current observation:
quasi-degenerate~\cite{Ioannisian:1994nx,Babu:2002dz,Chankowski:2000fp},
normal, such as typical of seesaw models and bilinear R-parity
violation, and inverse-hierarchical neutrino masses.

Turning to the three mixing angles, they are a natural feature of
gauge theories and follow simply as a result of the fact that in
general I=1/2 (up-type) and I=-1/2 (down-type) Yukawa couplings (mass
matrices) are not simultaneously diagonal.  Typically mixing angles
are not predicted from first principles, as we lack a basic theory of
flavor. However there has been a flood of recent activity in trying to
\texttt{post-dict} neutrino mixing
angles~\cite{Babu:2002dz,Chankowski:2000fp,bi-max,Altarelli:gu}.

Finally, the simplest structure of the lepton mixing matrix implied by
a gauge theory of the weak interaction contains, in addition, three
$CP$ violating phases~\cite{Schechter:1980gr,Schechter:1980gk}.
\begin{itemize}
\item one Kobayashi-Maskawa-like $CP$ phase 
\item two Majorana-type $CP$ phases
\end{itemize}
The Majorana-type phases drop out from $\Delta L = 0$ processes, such
as standard oscillations~\cite{Schechter:1980gk,Doi:1980yb}.  As for
the ``Dirac'' $CP$ phase, it does appear in such lepton-number
conserving oscillations.  However, the corresponding $CP$ violation
disappears as two neutrinos become degenerate and/or as one of the
angles, e.~g.~$\theta_{13}$, is set to zero~\cite{Schechter:1979bn}.
Given the hierarchical nature of neutrino mass splittings, and the
smallness of the mixing angle $\theta_{13}$ indicated by reactor
experiments (see Sec.~\ref{sec:chooz}) it follows that probing $CP$
violation effects in oscillation experiments will be a very demanding
challenge.  Therefore all such phases will be neglected in our
discussion of solar and atmospheric neutrino oscillations.

In addition to the Majorana phases, the theoretically expected
structure of leptonic weak interactions is substantially more complex
in theories where neutrino masses arise from the so-called
mechanism~\cite{seesaw79,seesaw80,seesawmajoron}.  This follows from
the fact that such models contain \21 singlet leptons, so that the
full charged current (CC) mixing matrix is rectangular, and the
corresponding neutral current (NC) is
non-trivial~\cite{Schechter:1980gr}. In other words, the weak CC and
NC interactions of neutrinos becomes non-standard.
This implies yet additional angles and phases, which may lead to
lepton flavor violation, and leptonic $CP$ violation \texttt{even in
  the limit where neutrino masses would vanish}~\cite{Branco:bn}. This
has the important implication that such processes are unrestricted by
the smallness of neutrino mass.  Given the many possible variants of
the see-saw schemes~\cite{fae}, one finds that in some of such models
the iso-singlet leptons need not be super-heavy~\cite{NSImodels2},
their masses lying at the weak scale or so. This leads to sizeable
rates for lepton flavor and leptonic $CP$ violating processes,
unrelated to the magnitude of neutrino masses~\cite{Branco:bn}.

Insofar as neutrino propagation is concerned, we note that in this
class of models the \texttt{effective} CC neutrino mixing matrix is
not unitary, with a non-trivial neutrino mixing even in the massless
limit~\cite{first-NSI-resonance-paper}.
This brings in the possibility of resonant oscillations of massless
neutrinos in matter, first noted in~\cite{first-NSI-resonance-paper}.
Effectively, neutrino propagation in matter is non-standard, as
discussed in Sec.~\ref{sec:non-stand-inter}. For the time being we
neglect all these subtle features in the description of neutrino
oscillations we give in Sec.~\ref{sec:neutr-oscill}.

\begin{figure}[htbp] \centering
\includegraphics[width=0.7\linewidth,height=4.5cm]{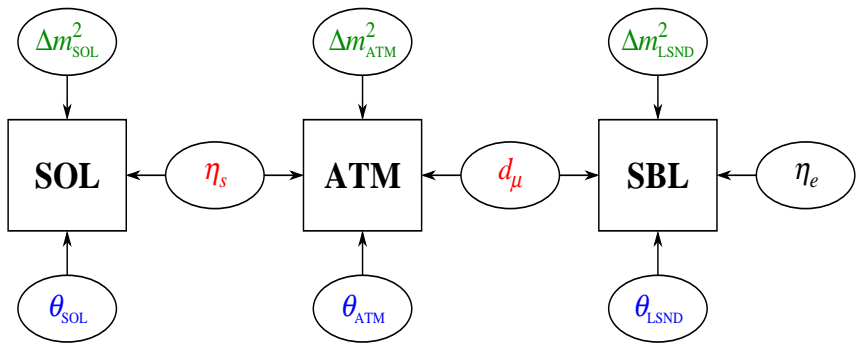}
\caption{Convenient separation of parameter dependence of the different data 
  sets used in Refs.~\cite{Maltoni:2001bc,Maltoni:2002xd}.}
\label{fig:diagram}
\vspace*{-2mm}
\end{figure}

It may also happen that some of the \21 singlet leptons are forced,
e.~g.~by a protecting symmetry, to remain light enough to participate
in the oscillations as \texttt{sterile
  neutrinos}~\cite{Peltoniemi:1992ss}.
Indeed, while the simplest three-neutrino picture is consistent with
all other oscillation searches, it fails to account for the LSND
hint~\cite{LSND}. Inclusion of the latter requires, in the framework
of the oscillation hypothesis, the existence of a fourth light sterile
neutrino taking part in the oscillations~\cite{Peltoniemi:1992ss}.
The presence of sterile neutrinos in the oscillations adds another
mass parameter, and also increases the number of mixing parameters to
six, in addition to $CP$ phases. 
A detailed parametrization was first given in
Ref.~\cite{Schechter:1980gr}. A simple factorization convenient for
use in a global analysis of oscillation data is illustrated in
Fig.~\ref{fig:diagram}.
%

%%%%%%%%%%%%%%%%%%%%%%%%%%%%%%

We will adopt this generalized framework in the description of solar
and atmospheric oscillations~\cite{Maltoni:2002ni} presented in
Secs.~\ref{sec:solar-neutr} and \ref{sec:atmosph-neutr} where we
describe, in particular, the constraints implied by both solar and
atmospheric data samples on the sterile admixture, $\eta_s$. On the
other hand this parametrization will also be employed in the global
analysis~\cite{Maltoni:2002xd} of all current oscillation data
presented in Sec.~\ref{sec:combining-lsnd-with}.

\subsection{The Absolute Scale of Neutrino Mass}
\label{sec:absol-scale-neutr}

Neutrino oscillations are sensitive only to mass splittings, not to
the absolute scale of neutrino mass.  Probing the latter requires
either direct kinematical tests, using tritium beta
spectrometers~\cite{katrin}, or observations of the Cosmic Microwave
Background and large scale structure, sensitive to a sub-leading hot
dark matter component~\cite{Elgaroy:2002bi}.  The present limits come
from a long list of painstaking efforts to study neutrino mass effects
in beta decays, which culminated with the Mainz and Troitsk results
(see Ref.  \cite{katrin} for the corresponding references).
Given the smallness of the solar and atmospheric mass splittings, the
resulting (conservative) bounds on the sum of all neutrino
masses~\cite{Elgaroy:2002bi,pdg} are illustrated in the ordinate of
Fig.~\ref{fig:bbmass}.

\begin{figure}[htbp] \centering
\includegraphics[width=0.4\textwidth,height=4cm]{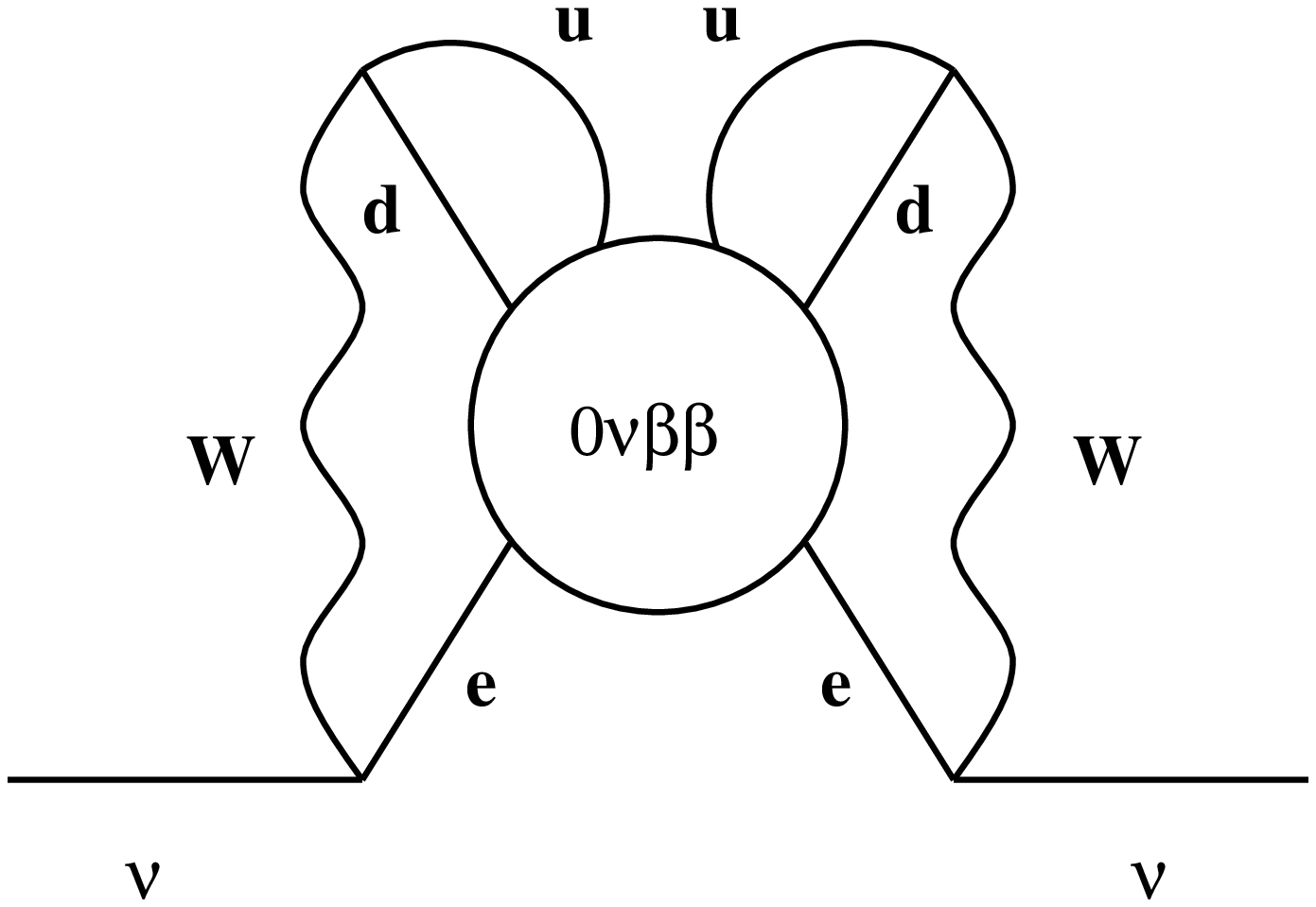}  
 \caption{The black-box \nbb argument \cite{Schechter:1981bd}.}
 \label{fig:bbox} 
\end{figure}

\begin{figure}[htbp] \centering
\includegraphics[width=0.45\textwidth,height=5cm]{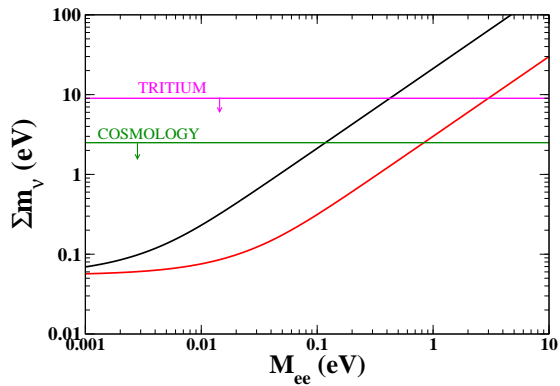}
  \caption{\nbb and the scale of neutrino mass.}
 \label{fig:bbmass} 
 \vspace*{-0mm}
\end{figure}

To decide whether neutrinos are Dirac or Majorana particles requires
the investigation of $\Delta L=2$ (L denoting lepton-number)
processes, of which \nbb decay provides the most classic example
\cite{Morales:1998hu}. Indeed, there is a \texttt{black-box
  theorem}~\cite{Schechter:1981bd} stating that, in a ``natural''
gauge theory, the observation of this process would signify the
discovery that neutrinos are, as expected by
theory~\cite{Schechter:1980gr}, Majorana fermions.  This connection is
illustrated by Fig.  \ref{fig:bbox}.
The importance of this simple argument lies in its generalality: it
holds irrespective of how \nbb is engendered. However, in order to
quantify its implications, one needs to specify the particular model.

In the neutrino-exchange-induced mechanism, \nbb is characterized by
an ``effective'' neutrino mass parameter $M_{ee}$ whose value is
sensitive to possible cancellations among individual neutrino
amplitudes. These may arise either as a result of
symmetries~\cite{pseudo,QDN} or due to the Majorana-type $CP$
phases~\cite{Schechter:1980gr}.
Nevertheless one can show that~\cite{Barger:2002xm}, as illustrated in
Fig.~\ref{fig:bbmass}, there is a direct correlation between $M_{ee}$
and the neutrino mass scales probed in tritium beta decays~\cite{pdg}
and cosmology~\cite{Elgaroy:2002bi}.  It is therefore important to
probe \nbb in a more sensitive
experiment~\cite{Klapdor-Kleingrothaus:1999hk}.
%

%%%%%%%%%%%%%%%%%%%%%%%%%%%%%%%%%%%%%%%%%%%%%%%%%%%%%%%%%%%%%%%%%%%%%%

\section{Neutrino Oscillations}
\label{sec:neutr-oscill}

Although the three-active neutrino oscillation scheme gives a good
description of both solar and atmospheric data, we will follow the
approach given in Ref.~\cite{Maltoni:2002ni} in which they are
analysed in terms of mixed active-sterile neutrino oscillations.
Such generalized scheme has as advantages that it allows one to
systematically combine solar and atmospheric data with the current
short baseline neutrino oscillation data samples including the LSND
evidence for oscillations~\cite{Maltoni:2002xd}, as done in
Sec.~\ref{sec:combining-lsnd-with}.
This is justified, since current reactor bounds on $\theta_{13}$
(Sec.~\ref{sec:chooz}) are stronger than solar and atmospheric bounds
on the parameter $\eta_s$ (with $0\le \eta_s \le 1$, see
Secs.~\ref{sec:solar-neutr} and \ref{sec:atmosph-neutr}) describing
the fraction of sterile neutrinos taking part in the solar
oscillations.
By taking such simplified analysis with $\theta_{13} \to 0$, we
completely decouple the solar and atmospheric oscillations from each
other, and comply trivially with the strong constraints from reactor
experiments.
For a complementary earlier analysis with $\theta_{13} \neq 0$
effects, but no sterile neutrinos, see Ref.~\cite{3-nu-sol+atm-fit}.
As seen in Fig.~\ref{fig:diagram}, mixed active-sterile neutrino
oscillations are characterized by a total of six mixing
angles~\cite{Schechter:1980gr}.

\subsection{Solar Neutrinos}
\label{sec:solar-neutr}

The solar neutrino data include the solar neutrino rates of the
chlorine experiment Homestake~\cite{Cleveland:nv} ($2.56 \pm 0.16 \pm
0.16$~SNU), the most recent result of the gallium experiments
SAGE~\cite{Abdurashitov:1999bv}~($70.8 ~^{+5.3}_{-5.2}
~^{+3.7}_{-3.2}$~SNU) and GALLEX/GNO~\cite{Altmann:2000ft} ($70.8 \pm
4.5 \pm 3.8$~SNU), as well as the 1496-days Super-Kamiokande data
sample~\cite{Fukuda:2002pe}. The latter are presented in the form of
44 bins (8 energy bins, 6 of which are further divided into 7 zenith
angle bins).
In addition to this, we have the latest results from SNO presented in
Refs.~\cite{Ahmad:2002jz}, in the form of 34 data bins (17 energy bins
for each day and night period). Therefore, in our statistical analysis
there are $3+44+34=81$ observables.

The most popular explanation of solar neutrino experiments is provided
by the neutrino oscillations hypothesis.  For generality we follow the
approach given in Ref.~\cite{Maltoni:2002ni} in which they are
analysed in terms of mixed active-sterile neutrino oscillations, where
the electron neutrino produced in the sun converts to $\nu_x$ (a
combination of $\nu_\mu$ and $\nu_\tau$) and a sterile neutrino $\nu_s
\:$: $ \nu_e \to \sqrt{1-\eta_s}\, \nu_x + \sqrt{\eta_s}\, \nu_s$.

In such framework  the solar neutrino data are fit with three
parameters $\Dms$, $\theta_\Sol$ and $\eta_s$.
The parameter $\eta_s$ with $0\le \eta_s \le 1$ describes the fraction
of sterile neutrinos taking part in the solar oscillations, so that
when $\eta_s \to 0$ one recovers the conventional active oscillation
case. The main motivation for adopting such generalized scenarios is
the possibility of combining the solar and atmospheric data with short
baseline oscillations~\cite{Maltoni:2002xd}.  Four-neutrino mass
schemes~\cite{Peltoniemi:1992ss} are the most natural candidates to
accommodate solar and atmospheric mass-splittings with the hint from
LSND~\cite{LSND} indicating a large $\Delta m^2$, see
Sec.~\ref{sec:react-accel-neutr}.

\begin{figure*}[thbp] 
    \includegraphics[height=7.5cm,width=0.96\linewidth]{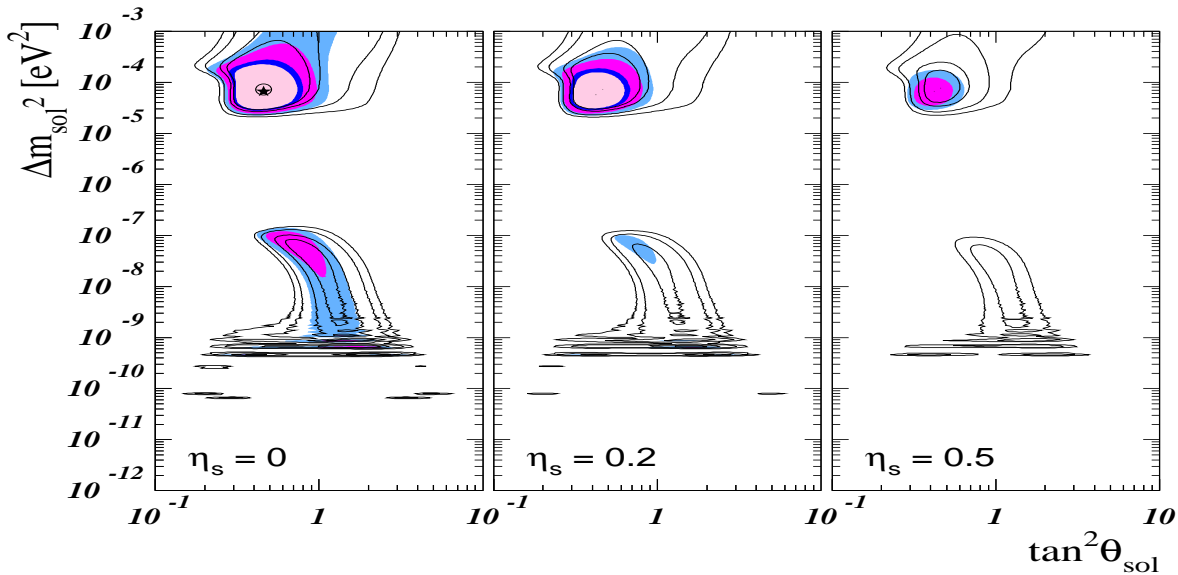}
\vspace*{-0mm}
      \caption{\label{fig:sol-osc-par}%
        Allowed $\tan^2\theta_\Sol$ and $\Dms$ regions for $\eta_s =
        0$ (active oscillations), $\eta_s = 0.2$ and $\eta_s = 0.5$.
        The lines and shaded regions correspond to the \snocc\ and
        \snotot analyses, respectively, as defined in
        Ref.~\cite{Maltoni:2002ni}.  The 90\%, 95\%, 99\% C.L. and
        3$\sigma$ contours are for 3 \dof.}
\vspace*{-0mm}
\end{figure*}

In Fig.~\ref{fig:sol-osc-par} we display the regions of solar neutrino
oscillation parameters for 3 \dof\ with respect to the global minimum,
for the standard case of active oscillations, $\eta_s = 0$, as well as
for $\eta_s = 0.2$ and $\eta_s = 0.5$.
The first thing to notice is the impact of the SNO NC, spectral, and
day-night data in improving the determination of the oscillation
parameters: the shaded regions after their inclusion are much smaller
than the hollow regions delimited by the corresponding \snocc\ 
confidence contours. Especially important is the full \snotot\ 
information in closing the LMA-MSW region from above: values of $\Dms
> 10^{-3}~\eVq$ appear only at $3\sigma$. Previous solar data on their
own could not close the LMA-MSW region, only the inclusion of reactor
data~\cite{CHOOZ} probed the upper part of the LMA-MSW
region~\cite{3-nu-sol+atm-fit}. Furthermore, the complete
\snotot\ information is important for excluding \texttt{maximal} solar
mixing in the LMA-MSW region.  At $3\sigma$ with 1 \dof\  one has
\begin{equation}\label{eq:sol_ranges}
    \rm{LMA-MSW:}\quad
    0.26 \le \tan^2\theta_\Sol \le 0.85 \,, \quad
    2.6\times 10^{-5}~\eVq \le \Dms \le
    3.3\times 10^{-4}~\eVq 
\end{equation}
showing that in the LMA-MSW region $\tan^2\theta_\Sol$ is significantly
below maximal.

%%%%%%%%%%%%

Note that in order to compare the allowed regions in
Fig.~\ref{fig:sol-osc-par} with others~\cite{Lisi}, one must note that
our \CL\ regions correspond to the 3 \dof\ corresponding to
$\tan^2\theta_\Sol$, $\Dms$ and $\eta_s$.  Therefore at a given \CL\ 
our regions are larger than the usual regions for 2 \dof, because we
also constrain the parameter $\eta_s$.
Our global best fit point occurs for active oscillations with 
\begin{equation}
\label{t12d12}
     \rm{LMA-MSW:}\quad \tan^2\theta_\Sol = 0.46\,, \quad
    \Dms = 6.6\times 10^{-5}~\eVq
\end{equation}

%%%%%%%%%%%%%%%%%%%%%%%%%%%%%%%%%%%%%%%%%%%%%%%%%%%%%%%%%%%%%%%%%%%%%%

\begin{figure*}[ht] \centering
\includegraphics[height=6cm,width=0.98\textwidth]{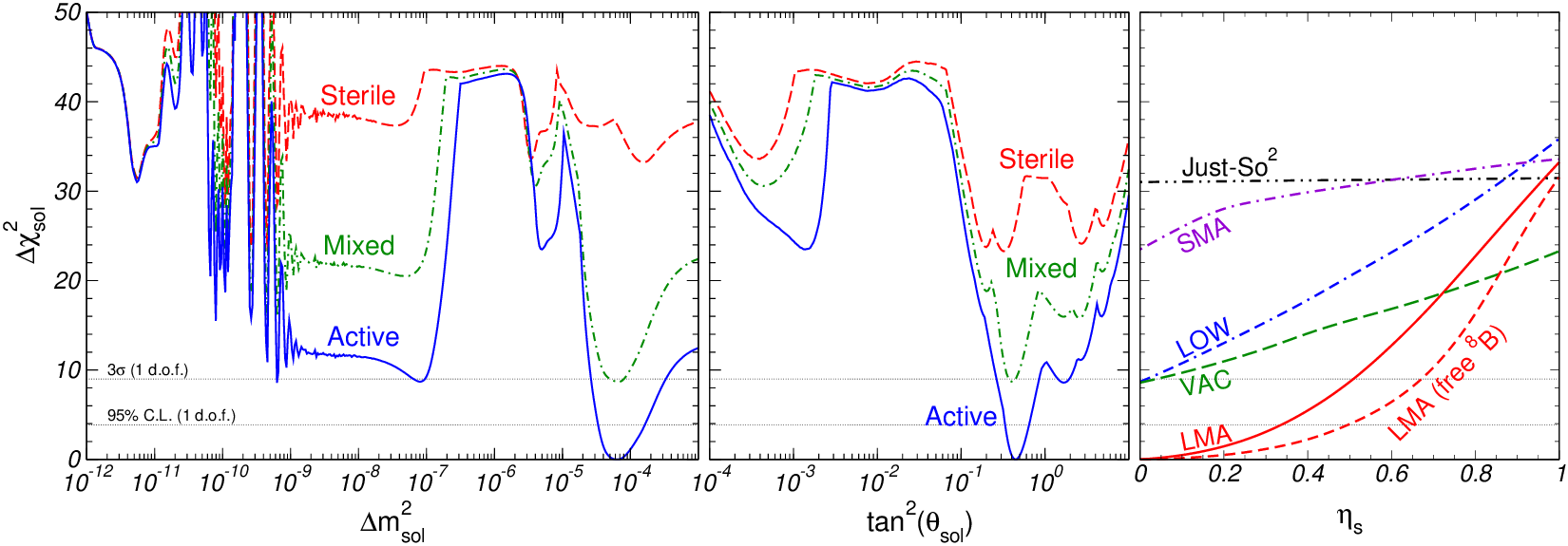}
\vspace*{-0mm}
  \caption{ \label{fig:chi-sol}%
    $\Delta\chi^2_\Sol$ as a function of $\Dms$, $\tan^2\theta_\Sol$,
    and $0 \leq \eta_s \leq 1$ from global \snotot\ sample defined in
    Ref.~\cite{Maltoni:2002ni}.}  \vspace*{-0mm}
\end{figure*}

A concise way to illustrate the above results is displayed in
Fig.~\ref{fig:chi-sol}. We give the profiles of $\Delta\chi^2_\Sol$ as
a function of $\Dms$ (left), $\tan^2\theta_\Sol$ (middle) as well as
$\eta_s$ (right), by minimizing with respect to the undisplayed
oscillation parameters.
In the left and middle panels the solid, dashed and dot-dashed lines
correspond to $\eta_s = 0$, $\eta_s = 1$ and $\eta_s = 0.5$,
respectively.
The use of the full \snotot\ sample has lead to the relative worsening
of all oscillation solutions with respect to the preferred active
LMA-MSW solution.
One sees also how the preferred status of the LMA-MSW solution
survives in the presence of a small sterile admixture characterized by
$\eta_s$. Increasing $\eta_s$ leads to a deterioration of all
oscillation solutions.
Note that in the right panel we display the profile of
$\Delta\chi^2_\Sol$ as a function of $0 \leq \eta_s \leq 1$,
irrespective of the detailed values of the solar neutrino oscillation
parameters $\Dms$ and $\theta_\Sol$.
One can see that there is a crossing between the LMA-MSW and VAC
solutions. This implies that the best pure--sterile description lies
in the VAC regime. However, in the global analysis pure sterile
oscillations with $\eta_s=1$ are highly disfavored. 
The $\chi^2$-difference between pure active and sterile is
$\Delta\chi^2_\mathrm{s-a} = 32.2$ if one restricts to the LMA-MSW
solution, or $\Delta\chi^2_\mathrm{s-a} = 23.3$ if one also allows for
VAC. For 3 \dof\ the $\Delta\chi^2_\mathrm{s-a} = 23.3$ implies that
pure sterile oscillations are ruled out at 99.997\% \CL\ compared to
the active case.

For the LMA-MSW solution one can also perform an analysis without
fixing the boron flux to its SSM prediction, as seen in the right
panel of Fig.~\ref{fig:chi-sol}. One can see that in this case the
constraint on $\eta_s$ is weaker than in the boron-fixed case, since a
{\it small} sterile component can now be partially compensated by
increasing the total boron flux coming from the Sun. From the figure
one obtains the bounds
\begin{equation}
    \label{eq:etasSol}
    \rm{solar  \: \: data:}
    \quad \eta_s \leq 0.44 \rm{~(boron-fixed)},
    \quad \eta_s \leq 0.61 \rm{~(boron-free)}
\end{equation}
at 99\% \CL\ for 1 \dof.
A complete table of best fit values of $\Dms$ and $\theta_\Sol$ with
the corresponding $\chi^2_\Sol$ and GOF values for pure active, pure
sterile, and mixed neutrino oscillations is given in
Ref.~\cite{Maltoni:2002ni}, both for the \snocc\ ($48-2$ \dof) and the
\snotot\ analysis ($81-2$ \dof).

{\bf Comparing different solar neutrino analyses}

Table \ref{tab:sol_comp} summarizes a compilation of the results of
the solar neutrino analyzes performed by the SNO and Super--K
collaborations, as well as by different theoretical groups (see
Ref.~\cite{Maltoni:2002ni} for the references).  All groups find the
best fit in the LMA-MSW region, although there are quantitative
differences even for this preferred solution. As can be seen from the
table, the GOF of the best-fit LMA-MSW solution, ranges from 53\% to
97\%.

Generally speaking, one expects the differences in the statistical
treatment of the data to have little impact on the global best fit
point, located in the LMA-MSW region.  These differences typically
become more visible as one compares absolute $\chi^2$ values or
departs from the best fit region towards more disfavored solutions.
Aware of this, Ref.~\cite{Maltoni:2002ni} took special care to details
such as the dependence of the theoretical errors on the oscillation
parameters entering the covariance matrix characterizing the Super-K
and SNO electron recoil spectra.  This ensures reliability of the
results in the full $\tan^2\theta_\Sol$-$\Dms$ plane.

The row labeled ``\dof''\ in table \ref{tab:sol_comp} gives the number
of analyzed data points minus the fitted parameters in each analysis.
We also present the best fit values of $\tan^2\theta_\Sol$ and $\Dms$
for active oscillations, the corresponding $\chi^2$-minima and GOF, as
well as the $\Delta\chi^2$ with respect to the favored active LMA-MSW
solution
\begin{table*}[!tbhp] 
    \catcode`?=\active \def?{\hphantom{0}}
 \centering \small
    \begin{tabular}{|>{\rule[-3mm]{0pt}{6mm}}l|c|c|c|c|c|c|c|c|c|c|c|}
        \hline
        & \rotatebox{90}{SNO Collaboration}
        & \rotatebox{90}{Super--K Collaboration}
        & \rotatebox{90}{Barger et al}
        & \rotatebox{90}{Bandyopadhyay et al}
        & \rotatebox{90}{Bahcall et al}
        & \rotatebox{90}{Creminelli et al}
        & \rotatebox{90}{Aliani et al}
        & \rotatebox{90}{De~Holanda \& Smirnov}
        & \rotatebox{90}{Fogli et al}
        & \rotatebox{90}{Barranco et al, Ref.~\cite{Barranco:2002te}}
        & \rotatebox{90}{Maltoni et al, Ref.~\cite{Maltoni:2002ni}} \\ 
        \hline
        \dof & 75-3 & 46 & 75-3 & 49-4 & 80-3 & 49-2 & 41-4 & 81-3 & 81-3 & 81-2 & 81-2 \\
        \hline
      \hline
        best OSC-fit & \multicolumn{11}{c|}{active LMA-MSW solution} \\
        \hline
        $\tan^2\theta_\Sol$
        & 0.34 & 0.38 & 0.39 & 0.41 & 0.45 & 0.45 & 0.40 & 0.41 & 0.42 & 0.47 & 0.46 \\
         $\Dms$ [$10^{-5}$ eV$^2$]
        & 5.0 & 6.9 & 5.6 & 6.1 & 5.8 & 7.9 & 5.4 & 6.1 & 5.8 & 5.6 & 6.6 \\
        $\chi^2_\mathrm{\bf LMA}$
        & 57.0 & 43.5 & 50.7 & 40.6 & 75.4 & 33.0 & 30.8& 65.2 & 73.4 &  68.0 &  65.8 \\
        GOF
        & 90\% & 58\% & 97\% & 66\% & 53\% & 94\% & 80\% & 85\% & 63\% &  81\% &  86\% \\
        \hline
        $\Delta \chi^2_{\rm{\bf LOW }}$
        & 10.7 & ?9.0 & ?9.2 & 10.0 & ?9.6 & ?8.1 & --& 12.4 & 10.0 & -- &  ?8.7 \\
        $\Delta \chi^2_{\rm{\bf VAC }}$                        
        & --   & 10.0 & 25.6 & 15.5 & 10.1 & 14.? & --& ?9.7 & ?7.8 & -- &  ?8.6 \\
        $\Delta \chi^2_{\rm{\bf SMA }}$                        
        & --   & 15.4 & 57.3 & 30.4 & 25.6 & 23.? & --& 34.5 & 23.5 & -- &  23.5 \\
         \hline
   \end{tabular} 
   \caption{\label{tab:sol_comp}%
Comparison of  different solar neutrino analyzes before KamLAND, 
from Ref.~\cite{Maltoni:2002ni}. See text.}
\end{table*} 
One can see from these numbers how various groups use different
experimental input data, in particular the spectral and zenith angle
information of Super--K and/or SNO. Despite differences in the
analyzes there is relatively good agreement on the best fit active
LMA-MSW parameters: the best fit values for $\tan^2\theta_\Sol$ are in
the range $0.34- 0.47$ and for $\Dms$ they lie in the range $(5.0 -
7.9)\times 10^{-5}$ eV$^2$. There is also good agreement on the
allowed ranges of the oscillation parameters (not shown in the table).
For example, the $3\sigma$ intervals given in Bahcall et al ($0.24 \le
\tan^2\theta_\Sol \le 0.89,\: 2.3\times 10^{-5}\:\rm{eV}^2 \le \Dms
\le 3.7\times 10^{-4}\:\rm{eV}^2$) and Holanda-Smirnov
($\tan^2\theta_\Sol \le 0.84,\: 2.3\times 10^{-5}\:\rm{eV}^2 \le \Dms
\le 3.6\times 10^{-4}\:\rm{eV}^2$) agree very well with those given in
Ref.~\cite{Maltoni:2002ni}.
There is remarkable agreement on the rejection of the LOW solution
with respect to LMA-MSW with a $\Delta\chi^2_{\rm{LOW, active}}
\approx 10$. The result for the vacuum solution in
\cite{Maltoni:2002ni} $\Delta\chi^2_{\rm{VAC, active}} = 8.6$ is in
good agreement with the values obtained by the Super-K collaboration,
as well as Bahcall et al, de Holanda \& Smirnov and Fogli et al,
whereas Bandyopadhyay et al, Barger et al and Creminelli et al obtain
higher values.  For the SMA-MSW solution one finds $\Delta\chi^2_{\rm
  SMA, active} = 23.5$, in good agreement with the values obtained in
Bahcall et al, Creminelli et al and Fogli et al; while Bandyopadhyay
et al and de Holanda \& Smirnov, and especially Barger et al, obtain
higher values. Typically the results of a given analysis away from the
best fit LMA-MSW region serve as an indicator of its quality.

All in all, in view of the vast input data, of possible variations in
the choice of the $\chi^2$ function and the treatment of errors and
their correlations, and of the complexity of the codes involved, it is
encouraging that there is reasonable agreement amongst different
analyzes, especially for the LMA-MSW solution. As shown in
Sec.~\ref{sec:neutr-oscill-after}, LMA-MSW is now strongly preferred
after the results of KamLAND described in Sec.~\ref{sec:kamland-data}.
From this point of view it has now become somewhat academic to
scrutinize further the origin of the differences found in the various
analyses. Nature has chosen the simplest solution.

\subsection{Atmospheric Neutrinos}
\label{sec:atmosph-neutr}

Here we summarize the analysis of atmospheric data given in a
generalized oscillation scheme in which a light sterile neutrino takes
part in the oscillations~\cite{Maltoni:2002ni}.  For simplicity the
approximation $\Dms\ll\Dma$ is used and the electron neutrino is taken
as completely decoupled from atmospheric oscillations, by setting
$\theta_{13} \to 0$ (for an analysis with $\theta_{13} \neq 0$ see
\cite{3-nu-sol+atm-fit}).  This way we comply with the strong
constraints from reactor experiments in Sec~\ref{sec:chooz}.
In contrast with the case of solar oscillations, the constraints on
the $\nu_\mu$--content in atmospheric oscillations are not so
stringent.  Thus the description of atmospheric neutrino oscillations
in this general framework requires \texttt{two new parameters} besides
the standard two-neutrino oscillation parameters $\theta_\Atm$ and
$\Dma$.  The parameters $d_\mu$ and $d_s$ introduced in
Ref.~\cite{Maltoni:2001bc} and illustrated in Fig.~\ref{fig:diagram}
are defined in such a way that $1-d_\mu$ ($1-d_s$) corresponds to the
fraction of $\nu_\mu$ ($\nu_s$) participating in oscillations with
$\Dma$. Hence, pure active atmospheric oscillations with $\Dma$ are
recovered when $d_\mu=0$ and $d_s=1$. In four-neutrino models there is
a mass-scheme-dependent relationship between $d_s$ and the solar
parameter $\eta_s$. For details see Ref.~\cite{Maltoni:2001bc}.

To get a feeling on the physical meaning of these two parameters, note
that for $d_\mu=0$ the $\nu_\mu$ oscillates with $\Dma$ to a linear
combination of $\nu_\tau$ and $\nu_s$ given as
$ \nu_\mu \to \sqrt{d_s} \,\nu_\tau + \sqrt{1-d_s} \,\nu_s \,.$ For
earlier pure active descriptions see, for example,
Refs.~\cite{Fornengo:2001pm,Fornengo:2000sr} and papers therein.

\begin{figure*}[tbph] \centering
  \includegraphics[width=0.96\textwidth,height=9cm]{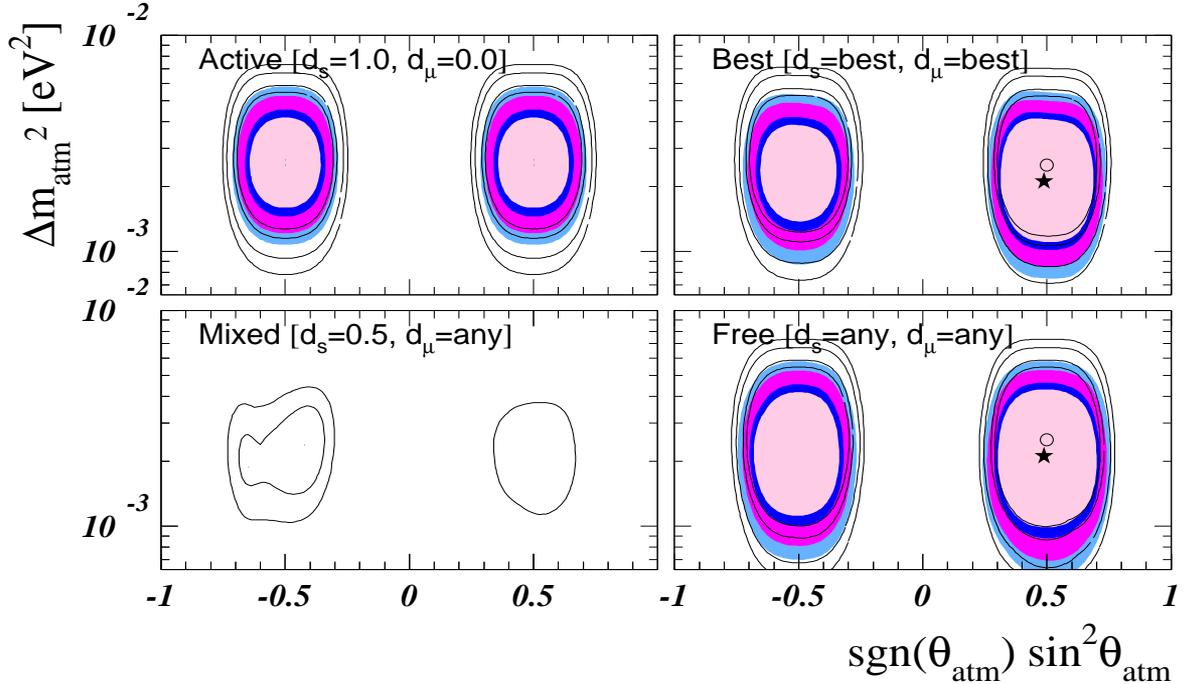}
\vspace*{-0mm}
  \caption{ \label{fig:atm-osc-par}%
    Allowed regions of $\sin^2\theta_\Atm$ and $\Dma$ at 90\%, 95\%,
    99\% and 3$\sigma$ for 4 \dof\ and different assumptions on the
    parameters $d_s$ and $d_\mu$, from~\cite{Maltoni:2002ni}. The
    lines (shaded regions) correspond to 1289 (1489) days of Super-K
    data. }  \vspace*{-0mm}
\end{figure*}

\begin{figure*}[bthp] \centering
  \includegraphics[width=\textwidth,height=5.5cm]{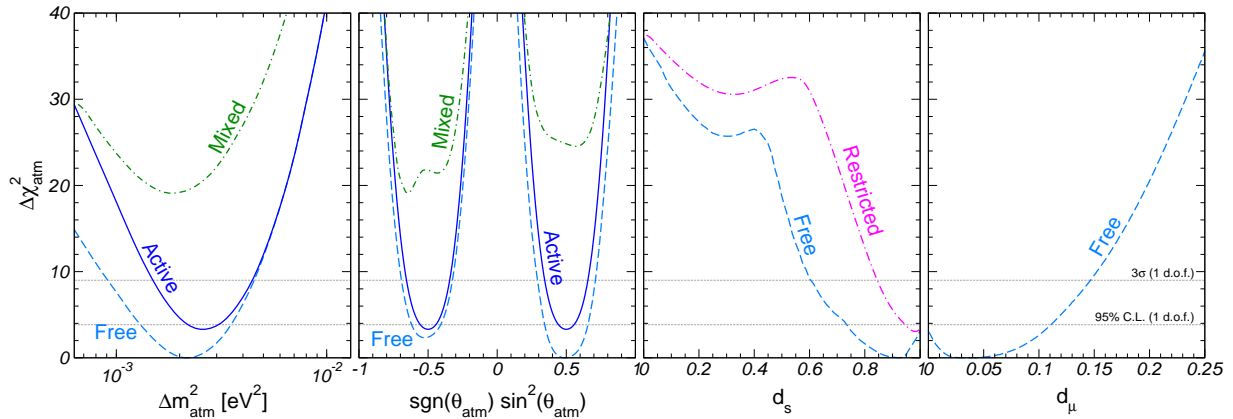}
  \vspace*{-0mm}
\caption{\label{fig:chi-atm}%
  $\Dcq_\Atm$ as a function of $\Dma$, $\sin^2\theta_\Atm$, $d_s$ and
  $d_\mu$. In each panel the undisplayed parameters are integrated
  out. The ``Mixed'' and ``Restricted'' cases correspond to $d_s =
  0.5$ and $d_\mu = 0$, respectively~\cite{Maltoni:2002ni}.}
\vspace*{-2mm} 
\end{figure*}

The global best fit point occurs at 
\begin{equation}
    \sin^2\theta_\Atm = 0.49 \,,\quad \Dma = 2.1 \times
    10^{-3}~\eVq 
\end{equation}
and has $d_s=0.92,\: d_\mu=0.04$. One sees that atmospheric data
prefers a small sterile neutrino admixture. However, this is not
statistically significant, since the pure active case ($d_s=1,
d_\mu=0$) also gives an excellent fit: the difference in $\chi^2$ with
respect to the best fit point is only $\Dcq_\mathrm{act-best} = 3.3$.
For the pure active best fit point one obtains,
\begin{equation}
\label{t23d23}
    \sin^2\theta_\Atm = 0.5 \,,\: \Dma = 2.5 \times
    10^{-3}~\eVq \: 
\end{equation}
with the 3$\sigma$ ranges (1 \dof)
\begin{eqnarray} 
    0.3 \le \sin^2\theta_\Atm \le 0.7 \\ 
    1.2 \times 10^{-3}~\eVq \le \Dma  \le 4.8 \times
    10^{-3}~\eVq \,. 
\end{eqnarray}
The determination of the parameters $\theta_\Atm$ and $\Dma$ is
summarized in Figs.~\ref{fig:atm-osc-par} and \ref{fig:chi-atm}.  
Note that Fig.~\ref{fig:chi-atm} considers several cases: arbitrary
$d_s$ and $d_\mu$, best--fit $d_s$ and $d_\mu$, and pure active and
mixed active--sterile neutrino oscillations, as indicated.

%%%%%%%%%%%%%%%%%%%%%%%%%%%%%%%%%%%%%%%%%%%%%%%%%%%%%%%%%%%%%%%%%%%%%%

At a given \CL\ the $\chi^2_\Atm$ is cut at a $\Dcq$ determined by 4
\dof\ to obtain 4-dimensional volumes in the parameter space of
($\theta_\Atm, \Dma, d_\mu,d_s$).  In the upper panels we show
sections of these volumes at values of $d_s=1$ and $d_\mu=0$
corresponding to the pure active case (left) and the best fit point
(right). Again one sees that moving from pure active to the best fit
does not change the fit significantly.  In the lower right panel both
$d_\mu$ and $d_s$ are projected away, whereas in the lower left panel
$d_s=0.5$ is fixed and one eliminates only $d_\mu$.
Comparing the regions resulting from 1489 days Super-K data (shaded
regions) with the one from the 1289 days Super-K sample (hollow
regions) we note that the new data leads to a slightly better
determination of $\theta_\Atm$ and $\Dma$. However, more importantly,
from the lower left panel we see how the new data show a much stronger
rejection against a sterile admixture: for $d_s=0.5$ no allowed region
appears at 3$\sigma$ for 4 \dof.

\begin{figure}[bthp] \centering
    \includegraphics[width=0.95\linewidth]{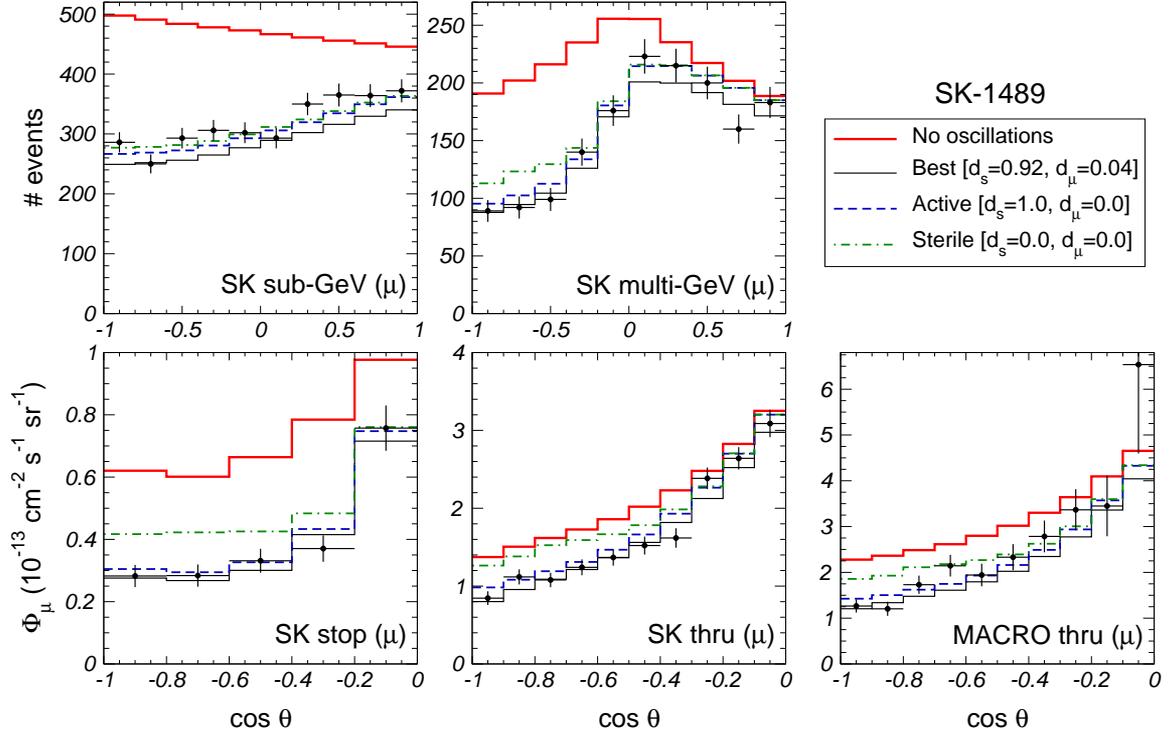}
    \caption{\label{fig:atm-zenith}%
      Zenith angle dependence of the $\mu$-like atmospheric neutrino
      data from Ref.~\cite{Maltoni:2002ni}.  The predicted number of
      atmospheric neutrino events for best--fit, pure--active and
      pure--sterile oscillations and no oscillations are given, for
      comparison.}
\end{figure}

The excellent quality of the neutrino oscillation description of the
present atmospheric neutrino data can be better appreciated by
displaying the zenith angle distribution of atmospheric neutrino
events, given in Fig.~\ref{fig:atm-zenith}.
Clearly, active neutrino oscillations describe the data very well
indeed. In contrast, the no-oscillations hypothesis can be visually
spotted as being ruled out. On the other hand, conversions to sterile
neutrinos lead to an excess of events for neutrinos crossing the core
of the Earth, in all the data samples except sub-GeV.

%%%%%%%%%%%%%%%%%%%%%%%%%%%%%%%%%%%%%%%%%%%%%%%%%%%%%%%%%%%%%%%%%%%%%%

\subsection{Reactor and Accelerator Neutrino Data}
\label{sec:react-accel-neutr}

\subsubsection{Chooz and Palo Verde}
\label{sec:chooz}

\begin{figure}[htbp]
  \centering
\includegraphics[width=.55\textwidth,height=6cm]{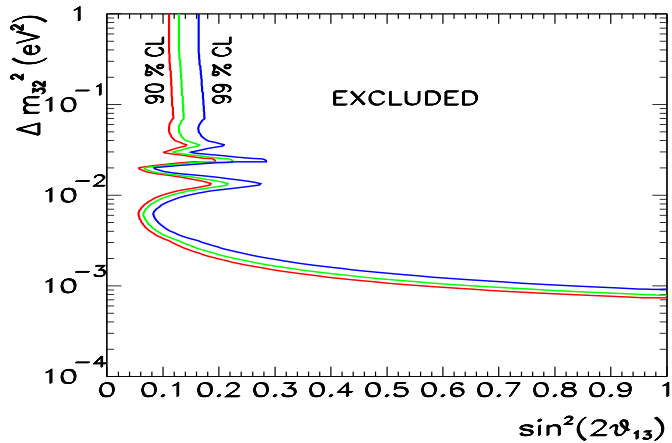}
  \caption{Region in  $\Delta{m}^2_{32}$ and $\sin^2(2\theta_{13})$  
  excluded by the Chooz reactor, from \cite{3-nu-sol+atm-fit}.}
  \label{fig:chooz}
\end{figure}

The Chooz experiment has been the first relatively long-baseline
reactor neutrino experiment.  As used in Ref.~\cite{Maltoni:2001bc},
the measured $\bar\nu_e$ survival probability from these experiments
are $P=1.01\pm 0.028\pm 0.027$ for Chooz, and $P =1.01\pm 0.024 \pm
0.053$ for Palo Verde~\cite{CHOOZ}.
The non-observation of oscillations at these reactors provides an
important restriction on $\Delta{m}^2_{32}$ and
$\sin^2(2\theta_{13})$, as illustrated in Fig.~\ref{fig:chooz}.
The curves represent the 90, 95 and 99\% CL excluded region defined
with 2 d.~o.~f. for comparison with the Chooz published results.
For large $\Delta{m}^2_{32}$ this gives a stringent limit on
$\sin^2(2\theta_{13})$, but not for low $\Delta{m}^2_{32}$ values.
Together with atmospheric data this implies that $\theta_{13}$ must be
rather small.  As will be seen below $\Dma \gg \Dms$, i.~e. one has a
somewhat hierarchical structure of neutrino mass \texttt{splittings}.

\subsubsection{KamLAND}
\label{sec:kamland-data}

In the KamLAND reactor neutrino experiment the target for the
$\bar{\nu}_e$ flux consists of a spherical transparent balloon filled
with 1000 tons of non-doped liquid scintillator. The anti-neutrinos
are detected via the inverse neutron $\beta$-decay
\begin{equation}
    \bar{\nu}_e+p \to e^{+}+n\,.
    \label{decay}
\end{equation} 
The spectral data are given in 13 bins of prompt energy above 2.6 MeV
in Fig.~5 of Ref.~\cite{:2002dm}.

There have been already several papers analysing the first results of
the KamLAND experiment, here we follow Ref.~\cite{Maltoni:2002aw}.
The KamLAND data are simulated by calculating the expected number of
events in each bin for given oscillation parameters as
\begin{equation}
\label{noe}
    N_i^\mathrm{th}(\Delta m^2,\theta) = f
    \int dE_\nu  \sigma(E_\nu) 
    \\
    \sum_j \phi_j(E_\nu)  P_j(E_\nu,\Delta m^2,\theta) 
    \int_i dE_e R(E_e,E'_e)\,.
\end{equation}
Here $R(E_e, E_e')$ is the energy resolution function and $E_e, E_e'$
are the observed and the true positron energy, respectively, and an
energy resolution of $7.5\%/\sqrt{E(\mathrm{MeV})}$ is
assumed~\cite{:2002dm}.  The neutrino energy is related to the
positron energy by $E_\nu=E_e'+\Delta$, where $\Delta$ is the
neutron-proton mass difference. The integration interval over $E_e$ is
determined by the prompt energy interval in each bin. The neutrino
spectrum $\phi(E_\nu)$ from nuclear reactors is well known, the
phenomenological parameterization given in
Refs.~\cite{Vogel:iv,Murayama:2000iq} has been used. The average fuel
composition for the nuclear reactors given in Ref.~\cite{:2002dm} is
adopted and possible effects due to time variations in the fuel
composition have been neglected~\cite{Murayama:2000iq}.  The sum over
$j$ in Eq.~(\ref{noe}) runs over 16 nuclear plants, taking into
account the different distances from the detector and the power output
of each reactor (see Table~3 of Ref.~\cite{kamlandproposal}).  The
relevant detection cross section $\sigma(E_\nu)$ is given in
Ref.~\cite{Vogel:1999zy}. In the two-neutrino framework the
disappearance probability for the neutrinos coming from the reactor
$j$ is given by
\begin{equation}
    P_j(E_\nu,\Delta m^2,\theta) = 1 - 
    \sin^22\theta \sin^2\frac{\Delta m^2 L_j}{4E_\nu} \,.
\end{equation}
The normalization factor $f$ in Eq.~(\ref{noe}) is determined in such
a way that for the case of no oscillations the total number of events
is 86.8, as expected from the Monte-Carlo simulation used in
Ref.~\cite{:2002dm}.

For the statistical analysis one uses the $\chi^2$-function
\begin{equation}
    \chi^2 = \sum_{i,j} (N_i^\mathrm{th} - N_i^\mathrm{obs})
    S^{-1}_{ij} (N_j^\mathrm{th} - N_j^\mathrm{obs}) \,.
\end{equation}

\begin{figure}[htbp!] \centering
    \includegraphics[height=6cm,width=0.55\textwidth]{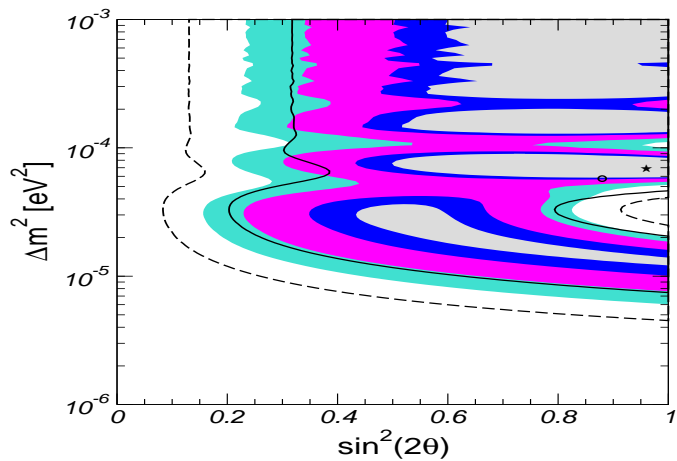}
    \caption{\label{fig:kamland}%
      Allowed regions at 90\%, 95\%, 99\% and 99.73\% \CL\ (2~\dof)
      from KamLAND spectral data. The solid (dashed) line is the 95\%
      \CL\ ($3~\sigma$) region from the KamLAND rate alone. The star
      (dot) is the best fit point from the spectral (rate) analysis,
      from Ref.~\cite{Maltoni:2002aw}.}
\end{figure}

The observed number of events $N_j^\mathrm{obs}$ in each bin can be
read off from Fig.~5 of Ref.~\cite{:2002dm}. In the covariance matrix
one includes the statistical errors (obtained from the same figure)
and the systematic error implied by the 6.42\% uncertainty on the
total number of events expected for no oscillations~\cite{:2002dm}.

In Fig.~\ref{fig:kamland} we show the allowed regions of the
oscillation parameters obtained from our re-analysis of the KamLAND
data. It is in good agreement with the analysis performed by the
KamLAND collaboration, shown in Fig.~6 of Ref.~\cite{:2002dm}. After
this successful calibration we turn to a full global analysis
combining also with the solar data sample of
Sec.~\ref{sec:neutr-oscill-after}.

\subsubsection{K2K}
\label{sec:k2k}

Further evidence for the atmospheric neutrino anomaly has now come
from the K2K experiment~\cite{Ahn:2002up} using accelerator neutrinos
in a long-baseline set-up.  The collaboration sees a reduction of the
$\nu_\mu$ flux together with a distortion of the energy spectrum. They
observe 56 beam neutrino events 250 km away from the neutrino
production point, with an expectation of $80.1^{+6.2}_{-5.4}$.  They
also reconstruct the neutrino energy spectrum, which fits better the
expected shape with neutrino oscillation than without.  The
probability that the observed flux at Super-K is a statistical
fluctuation without neutrino oscillation is less than 1\%.

\begin{figure}[htbp!] \centering
    \includegraphics[height=6cm,width=0.55\textwidth]{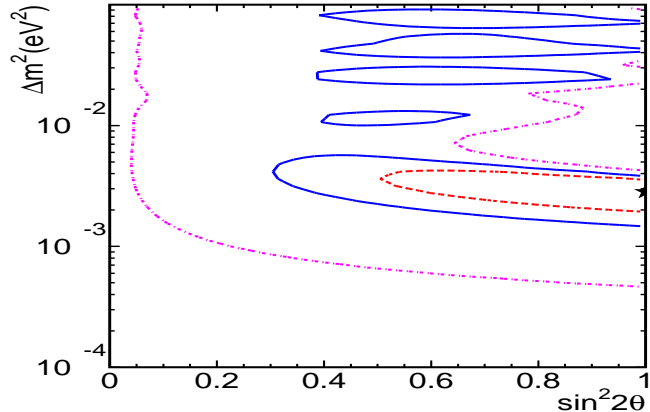}
    \caption{\label{fig:k2k}%
      Allowed regions of oscillation parameters from the K2K data of
      Ref.~\cite{Ahn:2002up}. Dashed, solid and dot-dashed lines are
      68.4\%, 90\% and 99\% C.L. contours, respectively.  The best fit
      point is indicated by the star.}
\end{figure}
The collaboration performs a two-neutrino oscillation analysis, with
\nm disappearance, using the maximum-likelihood method, and including
both the number of events and the energy spectrum shape. The results
are given in Fig.~\ref{fig:k2k} and agree nicely with what is inferred
from the atmospheric analysis, Sec.~\ref{sec:atmosph-neutr}.

\subsubsection{LSND }
\label{sec:lsnd}

The Liquid Scintillating Neutrino Detector (LSND) is an experiment
designed to search for neutrino oscillations in appearance channels.
It is the only short baseline accelerator neutrino experiment claiming
evidence for oscillations.

Here we compare the implications of two different analyses of the LSND
data.  The first uses the likelihood function obtained in the final
LSND analysis \cite{LSND} from their global data with an energy range
of $20 < E_e < 200$ MeV and no constraint on the likelihood ratio
$R_\gamma$ (see Ref.~\cite{LSND} for details). This sample contains
5697 events including decay-at-rest (DAR) $\bar\nu_\mu\to\bar\nu_e$,
and decay-in-flight (DIF) $\nu_\mu\to\nu_e$ data. We refer to this
analysis as \texttt{LSND global}. The second corresponds to the LSND
analysis performed in Ref.~\cite{Church:2002tc} based on 1032 events
obtained from the energy range $20 < E_e < 60$ MeV and applying a cut
of $R_\gamma > 10^{-5}$. These cuts eliminate most of the DIF events
from the sample, leaving mainly the DAR data, which are more sensitive
to the oscillation signal. We refer to this analysis as \texttt{LSND
  DAR}.

\begin{figure}[bthp]
  \centering 
  \includegraphics[width=0.6\linewidth,height=6.5cm]{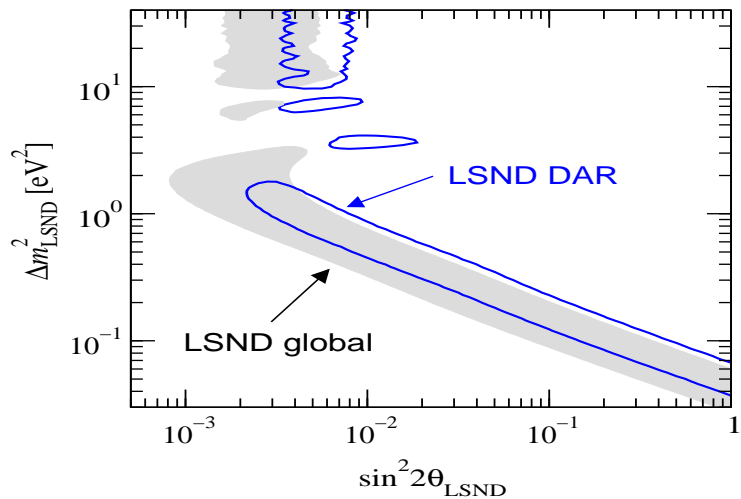}
  \caption{Comparison of the 99\% \CL\ regions of the global LSND
    analysis \cite{LSND} (shaded region) and of the analysis in
    Ref.~\cite{Church:2002tc} (solid line) using the decay-at-rest
    (DAR) sample. From Ref.~\cite{Maltoni:2002xd}}
  \label{fig:lsnd}
\end{figure}

In both cases the likelihood function obtained in the analyses of the
LSND collaboration was used and this was converted into a $\chi^2$
according to $\chi^2\propto -2\ln \mathcal{L}$ (see
Ref.~\cite{Maltoni:2001bc} for details).  In Fig.~\ref{fig:lsnd} we
compare the 99\% \CL\ regions obtained from the two LSND analyses. The
LSND DAR data prefers somewhat larger mixing angles, which will lead
to a stronger disagreement of the data in (3+1) oscillation schemes
(see below).  Furthermore, the differences in $\chi^2$ between the
best fit point and no oscillations for the two analyses are given by
$\Delta\chi^2_\mathrm{no\,osc} = 29$ (global) and
$\Delta\chi^2_\mathrm{no\,osc} = 47$ (DAR).  This shows that the
information leading to the positive oscillation signal seems to be
more condensed in the DAR data.  Note that the detailed information
from the short baseline disappearance no-evidence experiments
Bugey~\cite{bugey} and CDHS~\cite{CDHS} has been fully taken into
account.  Concerning the constraints from KARMEN~\cite{KARMEN}, they
are included by means of the KARMEN likelihood function.

\subsection{Neutrino Oscillations After KamLAND}
\label{sec:neutr-oscill-after}

\begin{figure}[ht] \centering
     \includegraphics[height=7cm,width=0.6\textwidth]{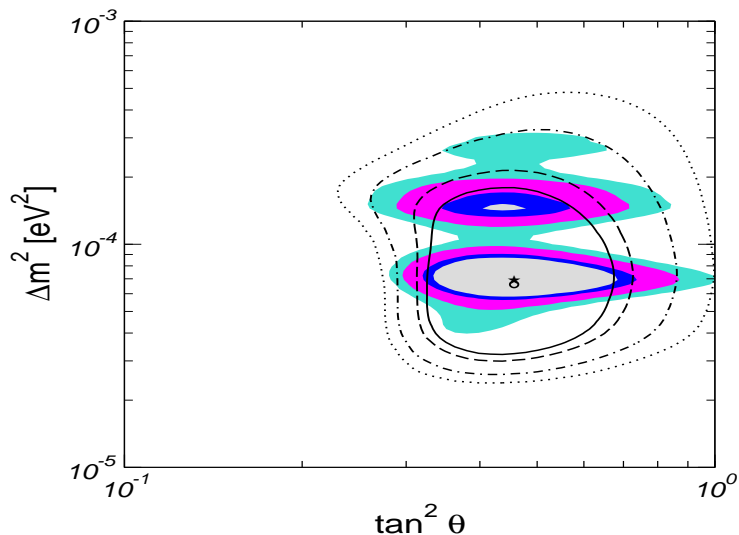}
    \caption{\label{fig:region}%
      Allowed regions at 90\%, 95\%, 99\% and 99.73\% \CL\ (2~\dof)
      from the combined analysis of solar, Chooz and KamLAND data. The
      hollow lines are the allowed regions from solar and Chooz data
      alone. The star (dot) is the best fit point from the combined
      (solar+Chooz only) analysis from Ref.~\cite{Maltoni:2002aw}.}
\end{figure}

\begin{figure*}[ht] \centering
    \includegraphics[width=0.95\textwidth,height=6cm]{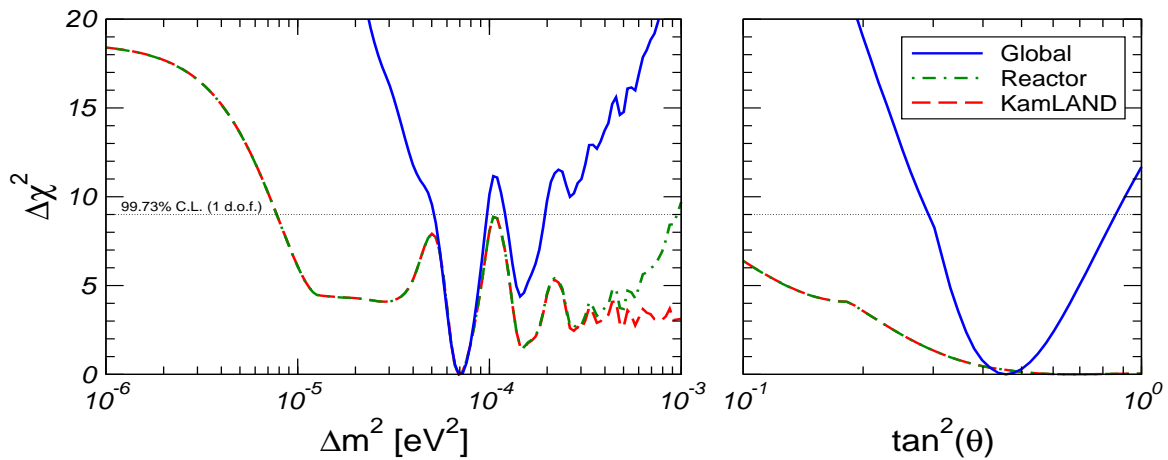}
    \caption{\label{fig:chisq}%
      $\Dcq$ versus $\Dms$ and $\tan^2 \theta$. The red dashed line
      refers to KamLAND alone. The green dot-dashed line corresponds
      to the full reactor data sample, including both KamLAND and
      Chooz. The blue solid line refers to the global analysis of the
      complete solar and reactor data, from
      Ref.~\cite{Maltoni:2002aw}.}
\end{figure*}

There has been a rush of recent papers on the analysis of neutrino
data after KamLAND in the framework of the neutrino oscillation
hypothesis (assuming, of course, $CPT$ invariance)
\cite{Maltoni:2002aw,kamland02others}.  Here we discuss the results of
the analysis presented in Ref.~\cite{Maltoni:2002aw}, to which the
reader is referred for the details.
Figs.~\ref{fig:region} and \ref{fig:chisq} summarize the results
obtained in a combined fit of the full KamLAND data sample with the
global sample of solar neutrino data (the same as used in
Ref.~\cite{Maltoni:2002ni}), as well as the Chooz result.

First of all, we have quantified the rejection of non-LMA solutions
and found that it is now more robust. For example, for the LOW
solution one has $\Dcq = 26.9$, which for 2~\dof\ ($\Dms$ and
$\theta$) lead to a relative probability of $1.4 \times 10^{-6}$. A
similar result is also found for the VAC solution. Besides selecting
out LMA-MSW as the unique solution of the solar neutrino problem we
find, however, that the new reactor results have little impact on the
location of the best fit point:
\begin{equation}
    \tan^2\theta = 0.46, \qquad \Dms = 6.9\times10^{-5}~\eVq.
\end{equation}
In particular the solar neutrino mixing remains significantly
non-maximal, a point which is not in conflict with the fact that
KamLAND data alone prefer maximal mixing~\cite{:2002dm}, since this
has no statistical significance~\cite{Maltoni:2002aw}. Indeed, one can
see from the right panel in Fig.~\ref{fig:chisq} that $\Dcq$ is rather
flat with respect to the mixing angle for $\tan^2 \theta \gsim 0.4$.
This explains why the addition of the KamLAND data has no impact
whatsoever in the determination of the solar neutrino oscillation
mixing. The allowed $3\sigma$ region one finds for $\theta$ is:
\begin{equation}
    0.29 \leq \tan^2\theta \leq 0.86,
\end{equation}
essentially the same as the pre-KamLAND range given in
Eq.~(\ref{eq:sol_ranges}).

Note that the solar mixing angle is large, but significantly
non-maximal, in contrast to the atmospheric mixing,
Eq.~(\ref{t23d23}). This important fact implies that models where the
solar mixing is non-maximal~\cite{Babu:2002dz} are strongly preferred
over bi-maximal mixing models~\cite{Chankowski:2000fp}.

Turning to the solar neutrino mass splittings, the new data do have a
strong impact in narrowing down the allowed $\Dms$ range. From the
left panel of Fig.~\ref{fig:chisq} one can see that the KamLAND data
alone provides the bound $\Dms > 8\times 10^{-6}~\eVq$, whereas the
CHOOZ experiment gives $\Dms < 10^{-3}~\eVq$, both at $3\sigma$. Hence
global reactor neutrino data provide a robust allowed $\Dms$ range,
based only on terrestrial experiments. However, combining this
information from reactors with the solar neutrino data leads to a
significant reduction of the allowed range: As clearly visible in
Fig.~\ref{fig:region}, the pre-KamLAND LMA-MSW region is now split
into two sub-regions. At $3\sigma$ (1 dof.) one obtains
\begin{equation}
    5.1\times 10^{-5}~\eVq \leq \Dms \leq 9.7\times 10^{-5}~\eVq, \\
    1.2\times 10^{-4}~\eVq \leq \Dms \leq 1.9\times 10^{-4}~\eVq.
\end{equation}
This remaining ambiguity might be resolved when more KamLAND data have
been collected~\cite{Murayama:2000iq,deGouvea:2001su,Barger:2000hy}.

\subsection{Combining LSND Data with the Rest}
\label{sec:combining-lsnd-with}

A possible confirmation of the LSND anomaly would have remarkable
implications. The most obvious would be the need for a sterile
neutrino, which should be light enough to participate in the
oscillations~\cite{Peltoniemi:1992ss}. There are two classes of
four-neutrino models, (3+1) and (2+2): in the first the sterile
neutrino can decouple from both solar and atmospheric oscillations,
while in the more symmetric (2+2) schemes, it can not decouple from
both sectors simultaneously~\cite{Maltoni:2001bc}. As a result (2+2)
schemes are now more severely rejected by a global
analysis~\footnote{Although KamLAND data are not included here,
  presently they have essentially no impact on the results presented
  in this section}.  
\begin{figure}[tbhp]
  \centering
  \includegraphics[width=0.75\linewidth,height=8cm]{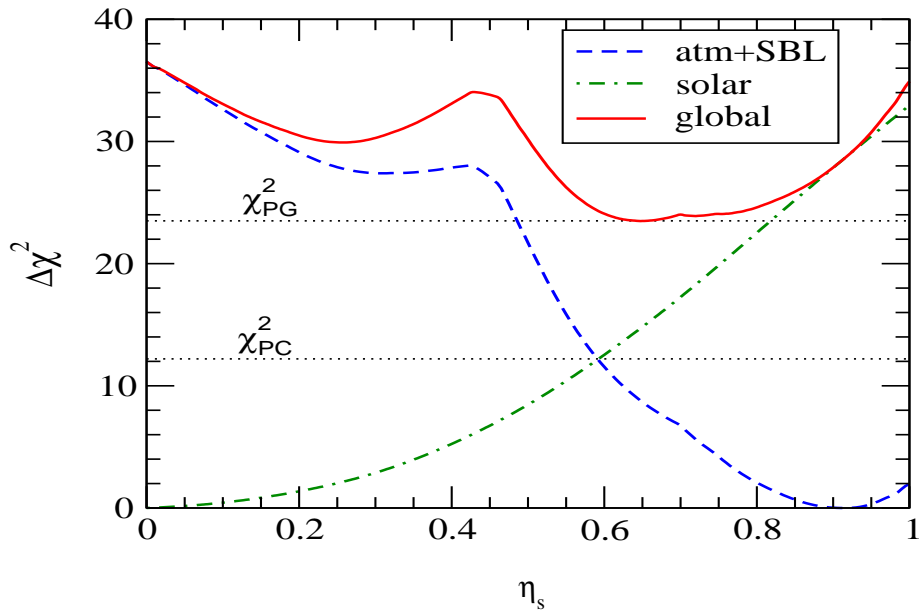}
  \caption{Excluding joint (2+2) oscillation descriptions of current 
    neutrino data including LSND, from Ref.~\cite{Maltoni:2002xd}.}
  \label{fig:sol+atm}
\end{figure}
Fig.~\ref{fig:sol+atm} shows the profiles of $\Delta\chi^2_\Sol$,
$\Delta\chi^2_{\Atm+\Sbl}$ and $\bar\chi^2_\mathrm{global}$ as a
function of $\eta_s$ in (2+2) oscillation schemes, as well as the
values $\chi^2_\mathrm{PC}$ and $\chi^2_\mathrm{PG}$ relevant for the
parameter consistency and parameter \gof\ tests proposed in
Ref.~\cite{Maltoni:2002xd}.  
\begin{figure}[htbp] %%\vglue -5cm
  \centering
  \includegraphics[width=0.75\linewidth,height=10cm]{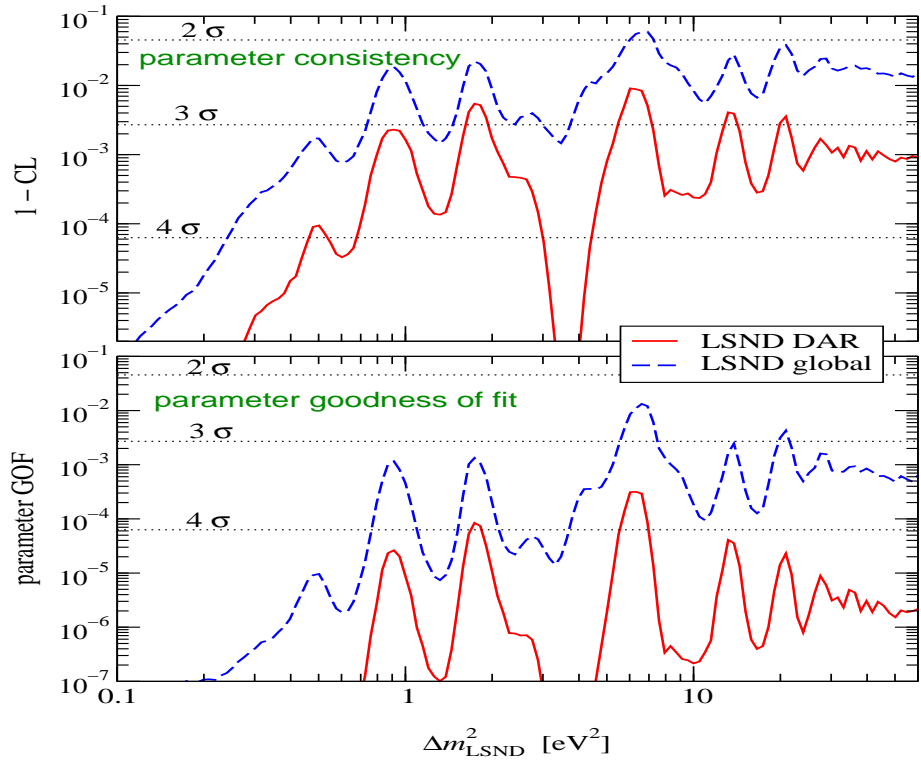}
  \caption{Compatibility of LSND with solar+atmospheric+NEV data in
    (3+1) schemes, from Ref.~\cite{Maltoni:2002xd}.}
  \label{fig:3+1}
\end{figure}
The application of these tests to quantify the compatibility of LSND
data with the remaining neutrino oscillation data in (3+1) schemes is
illustrated in Fig.\ref{fig:3+1}.  In the upper panel of
Fig.\ref{fig:3+1} we show the \CL\ of the parameter consistency,
whereas in the lower panel we show the parameter \gof\ for fixed
values of $\dml$. The analysis is performed both for the global
\cite{LSND} and for the DAR \cite{Church:2002tc} LSND data samples.
One sees that there is a slim chance to reconcile LSND data with the
remaining data, provided $\dml$ is close to 6 eV$^2$ or so, but only
at the expense of having a rather poor description.

In conclusion one finds that, though 4-neutrino models can not be
ruled out \texttt{per se}, the resulting global description of current
neutrino oscillation data is extremely poor, even in the case of (3+1)
schemes~\cite{Maltoni:2002xd}. We can only wait eagerly for news from
the upcoming MiniBooNE experiment.
Fortunately this experiment has began collecting data in the last
summer.  If it turns out that MiniBooNE ultimately confirms the LSND
claim we will face a real challenge.

\section{Neutrino Mixing in Cosmology and Astrophysics}
\label{sec:oscill-cosm}

Neutrino flavour mixing usually has no observational consequences in
Cosmology~\cite{Dolgov:2002wy} because in the standard cosmological
model all three neutrino flavours were produced in the early universe
with identical spectra, and thus with the same energy and number
densities.  However, it could be that any of the neutrino chemical
potentials was initially non-zero, or equivalently that a relic
asymmetry between neutrinos and antineutrinos existed, which in turn
increases the neutrino energy density and constitutes an extra
radiation density.  Only mild bounds on neutrino asymmetries exist
from the analysis of CMBR anisotropies, while Primordial Big Bang
Nucleosynthesis (BBN) places a more restrictive limit on the electron
neutrino chemical potential, because the $\bar{\nu}_{e}$ participates
directly in the beta processes that determine the primordial
neutron-to-proton ratio.

As seen in Sec.~\ref{sec:neutr-oscill-after}, the KamLAND data have
essentially fixed that neutrino oscillations explain the Solar
Neutrino Problem with parameters in the LMA-MSW region. It was shown
in Ref.~\cite{Dolgov:2002ab} that this result, combined with the
evidence of oscillations of atmospheric neutrinos, implies that
effective flavour equilibrium is established between all active
neutrino species before BBN. Therefore the BBN constraints on the
electron neutrino asymmetry apply to all flavours, which in turn
implies that neutrino asymmetries do not significantly contribute to
the extra relativistic degrees of freedom.  Thus the number density of
relic neutrinos is very close to its standard value, in such a way
that future measurements of the absolute neutrino mass scale, for
instance in the forthcoming tritium decay experiment
KATRIN~\cite{katrin}, will provide unambiguous information on the
cosmic mass density in neutrinos, free of the uncertainty of neutrino
chemical potentials.

If non-active light neutrino species exist, as suggested by the LSND
data, then they are resrticted also by BBN~\cite{Lisi:1999ng}.
However, this is less relevant now that the terrestrial data
themselves disfavor the light sterile neutrino oscillation hypothesis.

Back to three-neutrinos, the effect of large solar neutrino mixing in
astrophysics can be more substantial. First we note that the large
solar mixing angle opens the possibility of probing the noisy
character of the deep solar interior~\cite{Burgess:2002we}, especially
if an improved determination of $\Dms$ is available from further
KamLAND data.

Turning now to supernova neutrino spectra~\cite{Keil:2002in}, LMA-MSW
neutrino conversions in a supernova environment induce a significant
deformation of the energy spectra of neutrinos~\cite{Smirnov:ku}.
Despite this fact, a global analysis of SN1987A and solar neutrino
observations establishes the consistency of the LMA-MSW
solution~\cite{Kachelriess:2001sg}.  

\begin{figure*}[htbp]
  \centering
\includegraphics[width=.35\textwidth,height=3.8cm]{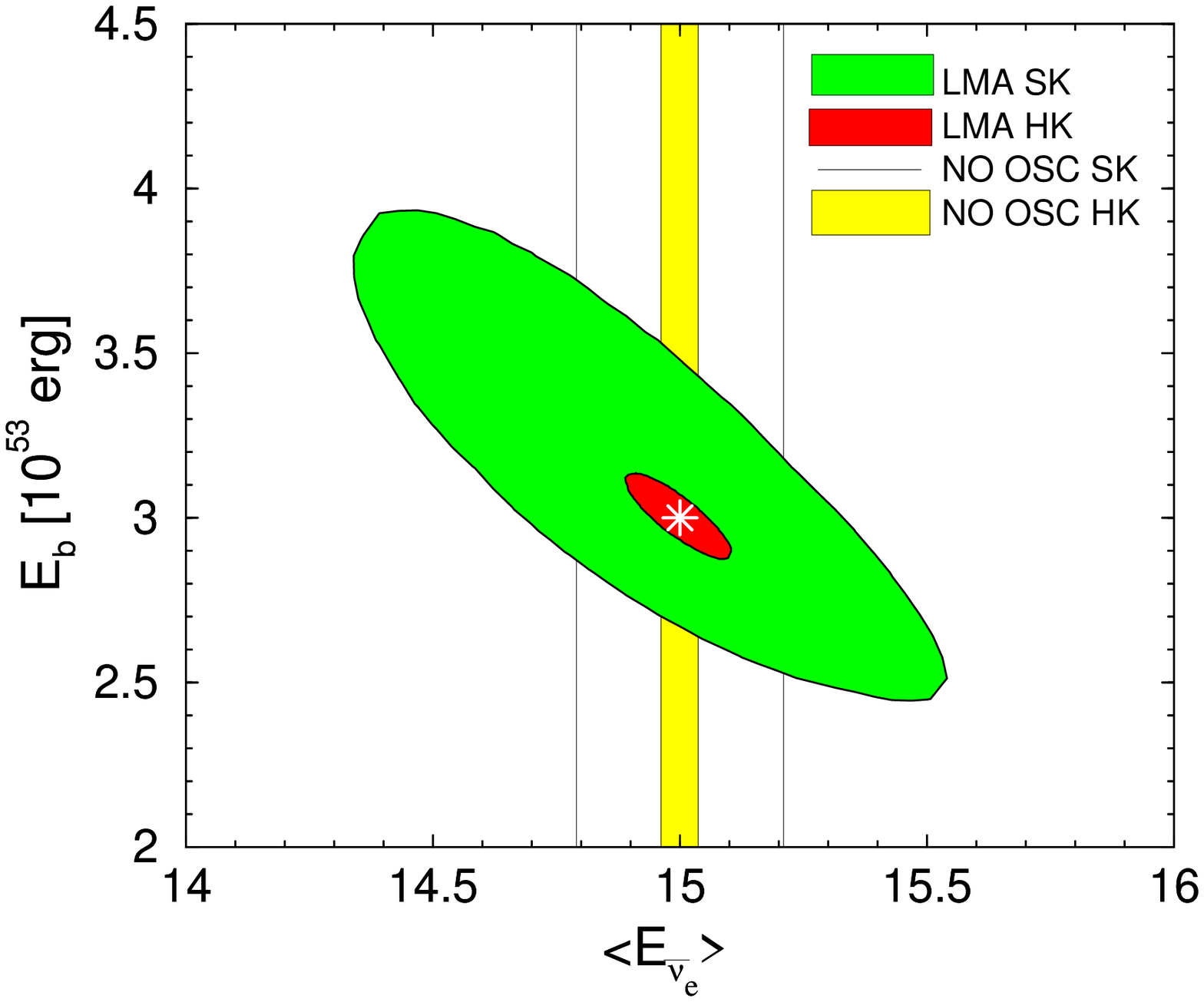}
\includegraphics[width=.35\textwidth,height=3.8cm]{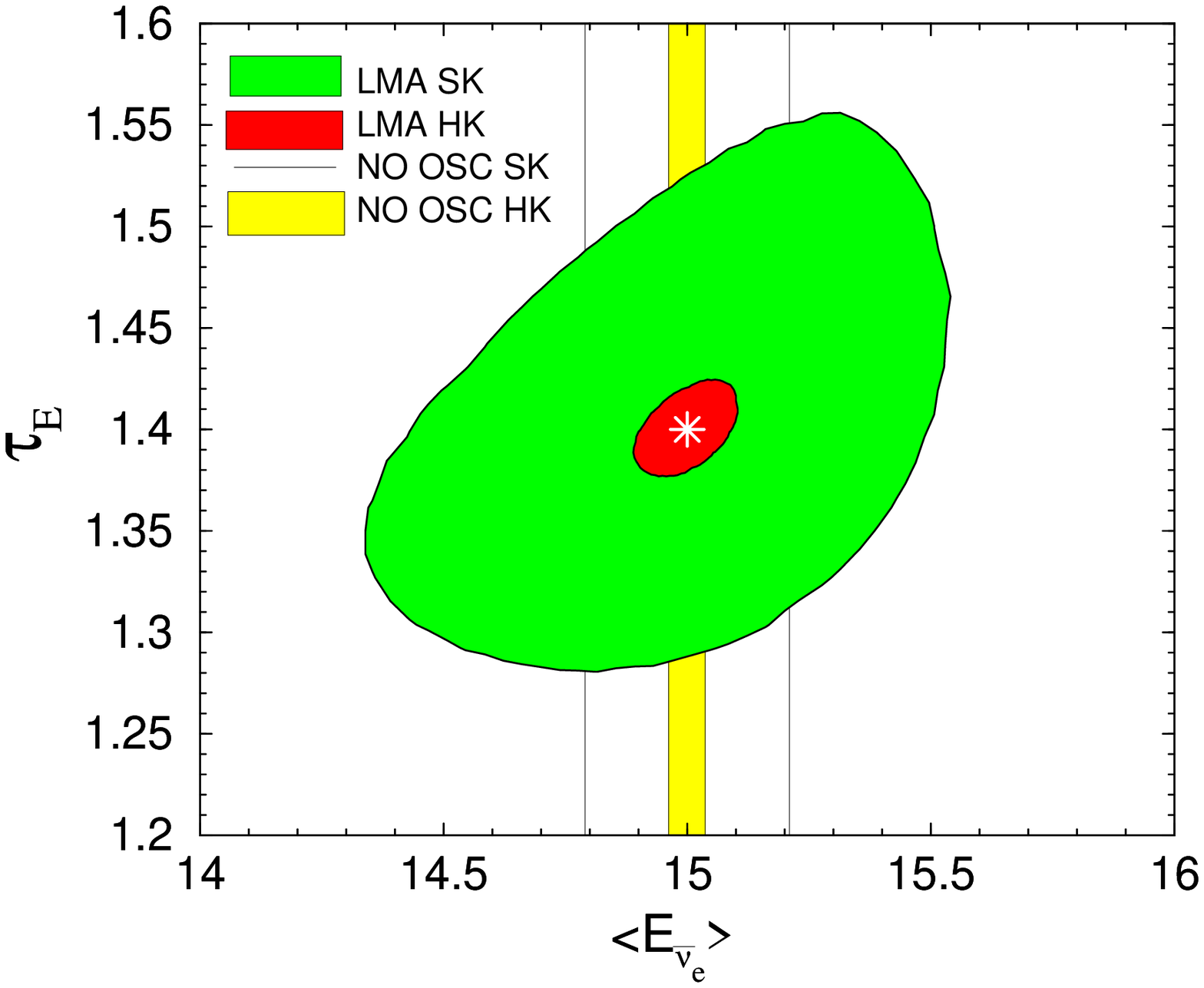}
\includegraphics[width=.35\textwidth,height=3.8cm]{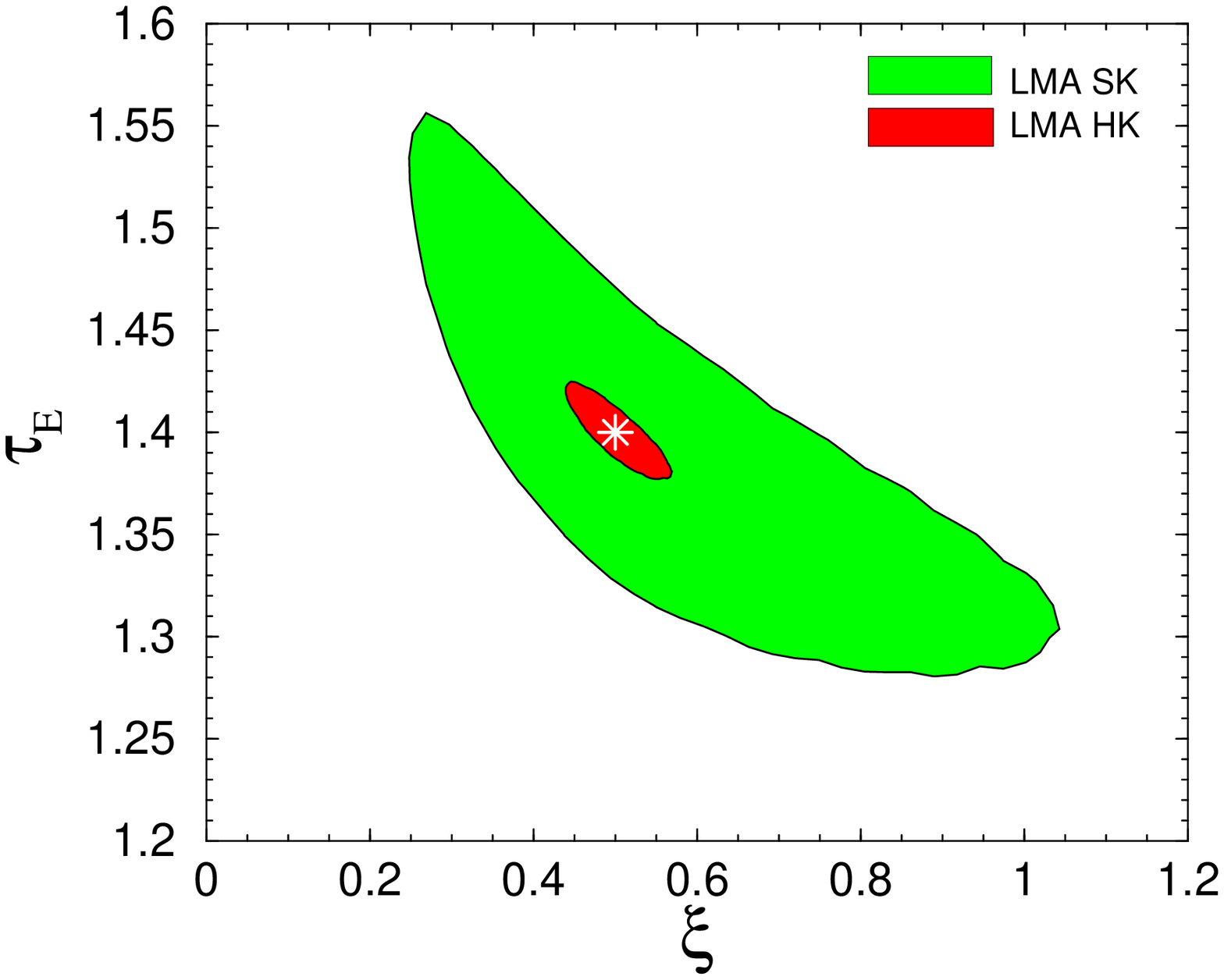}
\includegraphics[width=.35\textwidth,height=3.8cm]{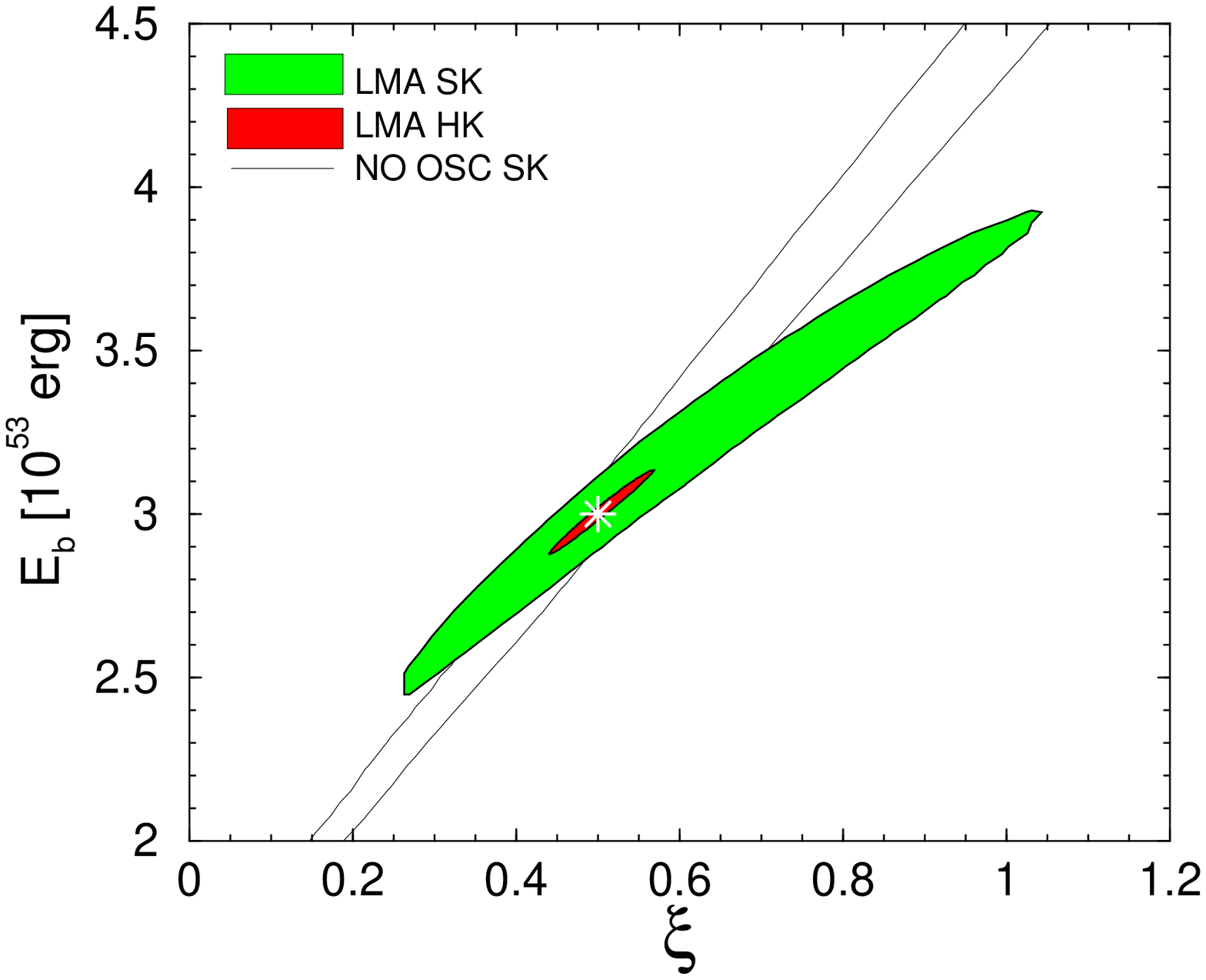}  
  \caption{Probing supernova spectra through LMA-MSW oscillations, 
    from Ref.~\cite{Minakata:2001cd}}
  \label{fig:minak}
\end{figure*}

Nevertheless, the large solar mixing angle does have a strong impact
on strategies for diagnosing collapse-driven supernovae through
neutrino observations, opening new ways to probe supernova parameters.
Indeed, fixing the LMA-MSW solution, one may in the future probe
otherwise inaccessible features of supernova neutrino spectra such as
the temperatures and luminosities of non-electron flavor
neutrinos~\cite{Minakata:2001cd}. This can be done simply by observing
$\bar{\nu}_{e}$'s from galactic supernovae through the charged current
reactions on protons, using massive water Cherenkov detectors.  As an
illustration we present Fig.~\ref{fig:minak} different 3 $\sigma$
contours for Super-K and Hyper-K, calculated both for the case of
LMA-MSW conversions and no-oscillations.  Best fits are indicated by
the stars. The plots result from a simulation which uses $\langle
E_{\bar{\nu}_e}^0 \rangle = 15$ MeV, $\tau^0$ = 1.4 and $E_b^0 = 3
\times 10^{53}$ erg as input supernova parameters. Details in
Ref.~\cite{Minakata:2001cd}.

Recent simulations indicate, however, that although the value of
$\tau^0$ may be substantially lower than what has been optimistically
assumed in Ref.~\cite{Minakata:2001cd}, the \texttt{fluxes} of
different flavors of supernova neutrinos may differ~\cite{Raffelt-pc},
giving an additional handle on the diagnostic of supernovae through
neutrino observations.

In addition to oscillations, other types of neutrino flavor
conversions can affect the propagation of neutrinos in a variety of
astrophysical environments, such as supernovae. For example, neutrino
non-standard interactions, discussed in
Sec.~\ref{sec:non-stand-inter}, could lead to resonant oscillations of
massless neutrinos in matter~\cite{first-NSI-resonance-paper}. These
could lead to ``deep-inside'' conversions~\cite{Nunokawa:1996tg},
rather distinct from those expected from conventional neutrino
oscillations~\cite{Fogli:2002xj}.

Another possibility are flavor conversions due to decays of neutrinos,
discussed in Sec.~\ref{sec:neutrino-decay}. If neutrino masses arise
from the spontaneous violation of ungauged lepton-number, they are
accompanied by a physical Goldstone boson, the
majoron~\cite{seesawmajoron}. In the high-density supernova medium the
effects of majoron-emitting neutrino decays are important even if they
are suppressed in vacuo by small neutrino masses and/or small
off-diagonal couplings.  Such strong enhancement is due to matter
effects, and implies that majoron-emitting decays have an important
effect on the neutrino signal of supernovae~\cite{Kachelriess:2000qc}.
majoron-neutrino coupling constants in the range $3\times 10^{-7}\lsim
g\lsim 2\times 10^{-5}$ or $g \gsim 3 \times 10^{-4}$ are excluded by
the observation of SN1987A, and could be probed with improved
sensitivity from a future galactic supernova.

\section{Non-standard Neutrino Interactions}
\label{sec:non-stand-inter}

Non-standard neutrinos interactions (NSI) are a natural feature in
most neutrino mass models~\cite{Schechter:1980gr,fae}. They
can be of two types: flavour-changing (FC) and non-universal (NU).
The co-existence of neutrino masses and NSI in many models of neutrino
mass means that neutrino flavor transformations may be induced by both
and therefore ideally both should be taken into account when analysing
neutrino data.

\begin{figure}[ht]
  \centering
\includegraphics[scale=.24]{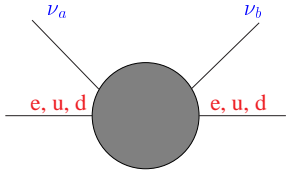}  
 \vspace*{-2mm}
 \caption{Effective  NSI operator.}
  \label{fig:nuNSI}
  \vspace*{-0mm}
\end{figure}

NSI may be schematically represented as effective dimension-6 terms of
the type $\varepsilon G_F$, as illustrated in Fig.~\ref{fig:nuNSI},
where $\varepsilon$ specifies their sub-weak strength.
Such interactions may arise from a nontrivial structure of CC and NC
weak interactions characterized a non-unitary lepton mixing matrix and
a correspondingly non-trivial NC matrix~\cite{Schechter:1980gr}.  Such
\texttt{gauge-induced} NSI may lead to flavor and $CP$ violation, even
with massless (degenerate) neutrinos \cite{NSImodels2}. In radiative
models where neutrino mass are ``calculable'' \cite{Zee:1980ai} and
supersymmetric models with broken R parity
\cite{Masiero:1990uj,rpmnu-exp} FC-NSI can also be
\texttt{Yukawa-induced}, from the exchange of spinless bosons.
In supersymmetric unified models, NSI may be calculable
renormalization effects~\cite{NSImodels3}.
We now describe the impact of non-standard neutrino interactions on
solar and atmospheric neutrinos. Since the NSI strengths are highly
model-dependent, we treat them as free phenomenological parameters.

\subsection{Solar Neutrinos}

At the moment one can not yet pin down the exact profile of the \ne
survival probability over the whole spectrum and, as a result, the
underlying mechanism of solar neutrino conversion remains unknown.
Thus non-oscillation solutions, such as those based on non-standard
neutrino matter interactions can be envisaged.
As already mentioned, an important feature of the NSI-induced
conversions~\cite{first-NSI-resonance-paper} is that the conversion
probability is energy-independent.
This implies that the solar neutrino energy spectrum is undistorted,
as indeed preferred by the Super-K and SNO spectrum data.

\begin{figure*}[ht]
  \centering
\includegraphics[width=0.96\textwidth,height=7cm]{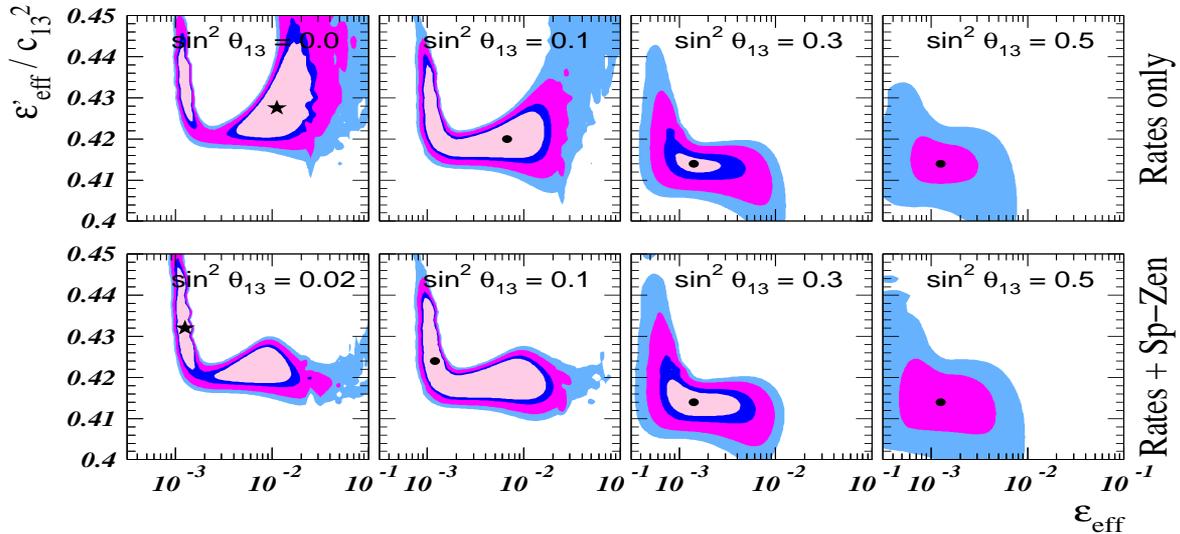}  
\vspace*{-2mm}
  \caption{Up-type quark NSI parameters needed to solve the solar 
    neutrino anomaly, from the first of Ref.~\cite{Guzzo:2001mi}.}
  \label{fig:hybparam}
\vspace*{-0mm}
\end{figure*}

In the first paper in Ref.~\cite{Guzzo:2001mi} it has been shown how
NSI provide an excellent description of present solar neutrino data.
The allowed regions for the NSI mechanism of solar neutrino conversion
are shown in Fig.~\ref{fig:hybparam}. Although the required magnitude
of NU interaction is somewhat large, it is not in conflict with current
data~\footnote{Note that there are no stringent direct bounds on NSI
  involving neutrinos, only for the charged leptons. However, the
  latter do not directly apply to the neutriunos case, hence the
  importance of the atmopsheric data}. 
Moreover one sees that the amount of FC interaction indicated by the
best fit is comfortably small.

Such a pure NSI description of solar data with massless and unmixed
neutrinos is slightly better than that of the favored LMA-MSW
solution, and the NSI values indicated by the solar data analysis do
not upset the successful oscillation description of the atmospheric
data~\cite{Guzzo:2001mi}. This establishes the overall consistency of
a hybrid scheme in which only atmopsheric data are explained in terms
of neutrino osciillations.
However, the recent first results of the KamLAND
collaboration~\cite{:2002dm} reject non-oscillation solutions, such as
those based on NSI, at more than 3~$\sigma$ so that the NSI effect in
solar neutrino propagation must be sub-leading. Accepting the LMA-MSW
solution one may determine restrictions on NSI parametetrs and,
therefore, on new aspects of neutrino mass models.

\subsection{Atmospheric Neutrinos}

\begin{figure}[tbh]
\centering
\includegraphics[width=0.45\textwidth,height=4.5cm]{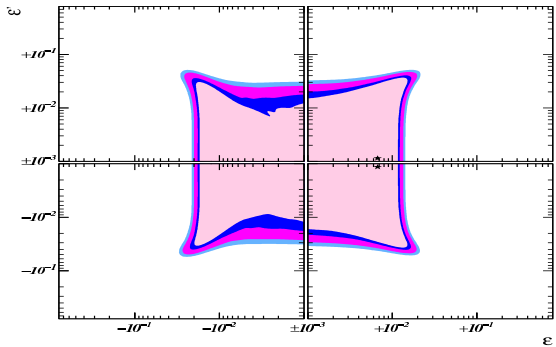}  
\vspace*{-2mm}
  \caption{Atmospheric bounds on neutrino NSI with 
    down-type quarks~\cite{Fornengo:2001pm}.}
  \label{fig:atmnsibds}
  \vspace*{-2mm}
\end{figure}

Flavor-changing non-standard interactions (FC-NSI) in the \nm-\nt
channel have been shown to account for the zenith--angle--dependent
deficit of atmospheric neutrinos observed in \texttt{contained}
Super-K events \cite{Gonzalez-Garcia:1998hj}.  The
solution works even in the absence of neutrino mass and mixing.
However such pure NSI explanation fails to reconcile these with
Super-K and MACRO \texttt{up-going muons}, due to the lack of energy
dependence intrinsic of NSI conversions. The discrepancy is at the
99\% 
\CL~\cite{Fornengo:2001pm}. Thus, unlike the case of solar
neutrinos, the oscillation interpretation of atmospheric data is
robust, NSI being allowed only at a sub-leading level.
Such robustness of the atmospheric \nm $\to$ \nt oscillation
hypothesis can be used to provide the most stringent current limits on
FC and NU neutrino interactions, as illustrated in
Fig.~\ref{fig:atmnsibds}.
These limits are rather model-independent, as they are obtained from
just neutrino-physics processes.  As described in
Sec.~\ref{sec:probing-nsi-with}, future neutrino factories can probing
non-standard neutrino interactions in this channel with better
sensitivity.

\section{Neutrino Magnetic Moments}
\label{sec:neutr-magn-moments}

\subsection{Intrinsic Magnetic Moments}
\label{sec:intr-magn}

\begin{figure}[bthp]
  \centering
  \includegraphics[width=0.5\linewidth,height=4.5cm]{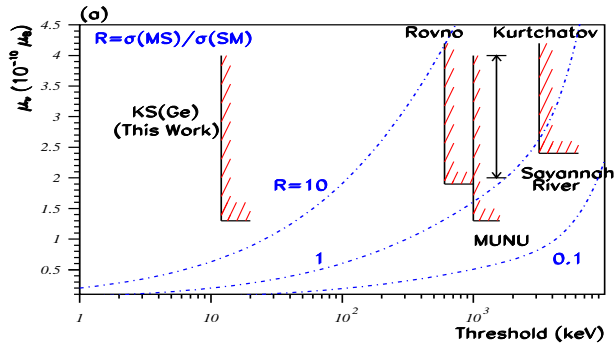}
  \caption{Summary of the results in 
    the searches of neutrino magnetic moments with reactor
    neutrinos, from Ref.~\cite{Li:2002pn}.}
  \label{fig:munu}
\end{figure}

Non-zero neutrino masses can manifest themselves through non-standard
neutrino electromagnetic properties.  When the lepton sector in the
Standard Model (SM) is minimally extended as in the quark sector,
neutrinos get Dirac masses ($m_\nu$) and their magnetic moments (MMs)
are tiny \cite{MMold},
\begin{equation}
  \label{eq:dirac}
\mu_\nu \simeq 3 \times 10^{-19}
\mu_B \left( \frac{m_\nu}{1 \, \mathrm{eV}} \right) \,,
\end{equation}
where $\mu_B$ is the Bohr magneton. Laboratory experiments give 90\%\CL\ 
bounds on the neutrino MMs of $1.8 \times 10^{-10} \mu_B$ \cite{rovno}
and $1.3 \times 10^{-10} \mu_B$ \cite{Li:2002pn,munu} for the electron
neutrino. These are summarized in~Fig.~\ref{fig:munu}.
For the muon neutrino the bound is $6.8 \times 10^{-10} \mu_B$ 
\cite{LSND2} and $3.9 \times 10^{-7} \mu_B$ for the tau neutrino
\cite{DONUT} (see also Ref.~\cite{pdg}). On the other hand,
astrophysics and cosmology provide limits of the order of $10^{-12}$
to $10^{-11}$ Bohr magnetons \cite{Raffelt:gv}. Improved sensitivity
for the electron neutrino from reactor neutrino searches is expected,
while a tritium $\bar\nu_e$ source experiment \cite{H3} aims to reach
the level $3 \times 10^{-12} \mu_B$.

It has for a long time been noticed, on quite general ``naturality''
grounds, that Majorana neutrinos constitute the typical outcome of
gauge theories~\cite{Schechter:1980gr}. On the other hand, precisely
such neutrinos also emerge in specific classes of unified theories, in
particular, in those employing the seesaw
mechanism~\cite{seesaw79,seesaw80,seesawmajoron}.  If neutrinos are
indeed Majorana particles the structure of their electromagnetic
properties differs crucially from that of Dirac
neutrinos~\cite{Schechter:1981hw}, being characterized by a $3 \times
3$ complex anti-symmetric matrix $\lambda$, the so-called Majorana
transition moment (\TM) matrix. It contains MMs as well as electric
dipole moments of the neutrinos.
The existence of any electromagnetic neutrino moment well above the
expectation in Eq.~(\ref{eq:dirac}) would signal the existence of
physics beyond the SM. Thus neutrino electromagnetic properties sre
sensitive probes of new physics. Majorana \TMs play an especially
interesting role. As we will describe next, they can affect neutrino
propagation in an important way and, to that extent, play an important
in cosmology and astrophsyics.

\subsection{Spin Flavor Precession}
\label{sec:spin-flav-prec}

Although LMA-MSW conversions are clearly favored over other
oscillation-type solutions, current solar neutrino data by themselves
are not enough to single out the mechanism of neutrino conversion
responsible for the suppression of the signal.

Magnetic-moment-induced neutrino conversions~\cite{MMold} in the
convective zone of the Sun~\cite{Okun:hi} have been long suggested as
a potential solution of the solar neutrino problem. However, this
would require too large neutrino magnetic moment and also that
neutrinos are Dirac particles, favored neither by
theory~\cite{seesaw79,seesaw80,seesawmajoron} nor by
astrophysics~\cite{Raffelt}. As a result here we focus on the
preferred case of Spin-flavor Precession
(SFP)~\cite{Schechter:1981hw,Akhmedov:uk}.

A global analysis of spin-flavour precession solutions to the solar
neutrino problem, taking into account the impact of the full set of
latest solar neutrino data, including the recent SNO-NC data as well
as the 1496--day Super-Kamiokande data has been given in
Ref.~\cite{Barranco:2002te}. These solutions depend in principle on
the magnetic field profile.  It is very convenient to adopt a
self-consistent form for the static magnetic field
profile~\cite{Miranda:2001bi,Miranda:2001hv} motivated by
magneto-hydrodynamics.  With this one finds that, to a good
approximation, the dependence of the neutrino SFP probabilities on the
magnetic field gets reduced to an effective parameter $\mu B_\perp$
characterizing the maximum magnetic field strength in the convective
zone. This way one is left with just three parameters: $\Dms \equiv
\Delta m^2$, the neutrino mixing angle $\theta_\Sol \equiv \theta$ and
the parameter $\mu B_\perp$. For $\mu = 10^{-11}$ Bohr magneton the
lowest optimum $B_\perp$ value is $\sim 80$ KGauss.

\begin{figure}[htbp] 
\centering 
\includegraphics[width=0.4\textwidth,height=5.5cm]{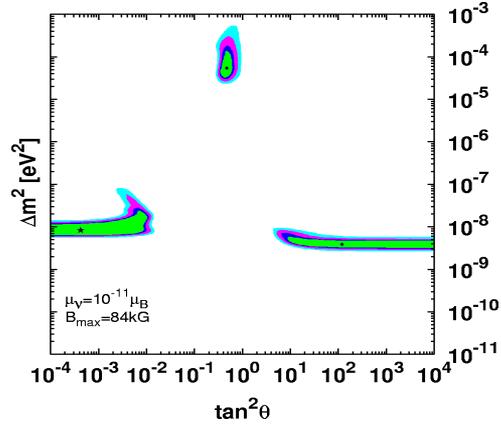}
 \vspace*{-2mm}
 \caption{Allowed $\Dms$ and $\tan^2\theta_\Sol$ for RSFP, LMA-MSW and NRSFP
   solutions for the indicated values of $\mu B$, from
   \cite{Barranco:2002te}}
 \label{sfpregions}
  \end{figure}

Fig.~\ref{sfpregions} shows the resulting parameter regions as given
in \cite{Barranco:2002te}.  One finds that, in addition to the
standard LMA-MSW solution, there are two SFP solutions, in the
resonant (RSFP) and non-resonant (NRSFP)
regimes~\cite{Miranda:2001hv}, with LOW-quasi-vacuum or vacuum
solutions absent at the 3 sigma level~\cite{Barranco:2002te}.
  Note that in the presence of a neutrino transition magnetic moment
  of $10^{-11}$ Bohr magneton, a solar magnetic field of 80 KGauss
  eliminates all oscillation solutions other than LMA-MSW,
  irrespective of KamLAND results.
\begin{figure}[tbhp] 
\centering 
\includegraphics[width=0.75\textwidth,height=4.5cm]{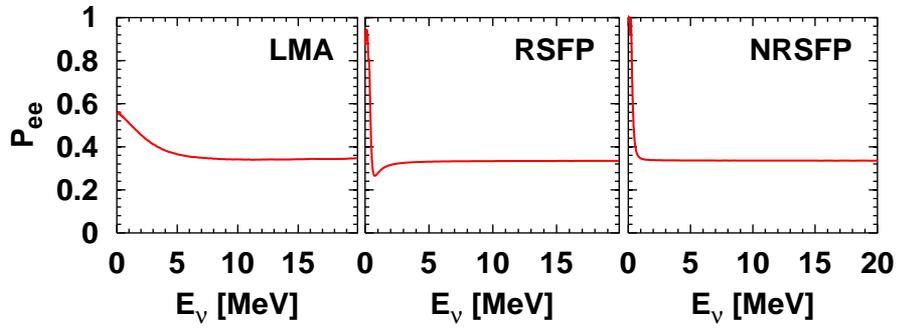}
\vspace*{-2mm}
\caption{Best LMA-MSW and SFP \ne survival probabilities from~\cite{Barranco:2002te}}
\label{probs.eps}
\vspace*{-2mm}
\end{figure}
  On the other hand Fig.~\ref{probs.eps} shows the predicted solar
  neutrino survival probabilities for the ``best'' LMA-MSW solution,
  and for the ``best'' SFP solutions, from latest solar data. Clearly
  the spectra in the high energy region are nearly undistorted in all
  three cases, in agreement with observations.
 \begin{figure}[thbp] 
\centering 
 \includegraphics[width=0.75\textwidth,height=4.5cm]{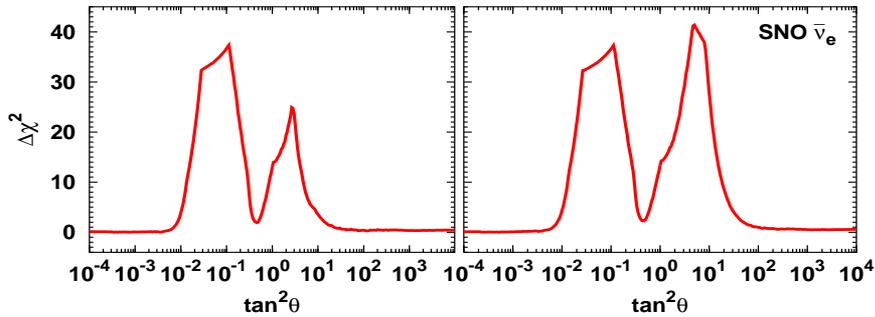}
 \vspace*{-0cm}
 \caption{$\Delta\chi^2_\Sol$ versus $\tan^2\theta_\Sol$, 
   for RSFP, LMA-MSW (central minima) and NRSFP solutions.  Left and
   right panels refer to two different analyses described in
   Ref.~\cite{Barranco:2002te}.}
 \label{chi2sfp}
 \vspace*{0mm}
 \end{figure}
 As far as the solar neutrino data are concerned, one finds that the
 two SFP solutions give a slightly lower $\chi^2_\Sol$ than LMA-MSW,
 though all three solutions are statistically equivalent.
 
 However, the recent first results announced by the KamLAND
 collaboration~\cite{:2002dm} imply that all non-oscillation solutions
 are strongly disfavored. For the case of SFP solutions one finds a
 rejection at about 3~$\sigma$, similar to that of non-LMA-MSW
 oscillation solutions before KamLAND.

\section{Neutrino Decay}
\label{sec:neutrino-decay}

It is generally agreed that most probably neutrinos have non-zero
masses and non-trivial mixings.  This belief is based primarily on the
evidence for neutrino mixings and oscillations from the solar and
atmospheric neutrino data.

If neutrinos are massive they can decay. Current information on the
absolute scale of neutrino mass from beta and double beta decay as
well as cosmology suggests neutrino masses are at most of order of eV.
Throughout the following discussion, we will stick to this assumption.
In this case the only neutrino decay modes available within the
simplest versions of the SM with massive neutrinos are radiative
decays of the type $\nu' \to \nu + \gamma$ and so-called invisible
decays such as the three-body decay $\nu' \to 3
\nu$~\cite{Schechter:1980gr,Schechter:1981cv} and two-body decays with
majoron emmission~\cite{seesawmajoron}.  The first one is
\texttt{``visible''}, while the latter two are \texttt{``invisible''}.

The questions are (a) whether the lifetimes are short enough to be
phenomenologically interesting and (b) what are the dominant decay
modes. The answer to these is, unfortunately, rather
model-dependent~\cite{fae}. 

\subsection{Radiative Decays}
\label{sec:radiative-decays}

For eV neutrinos, the only radiative decay modes possible are $\nu_i \to 
\nu_j + \gamma$.  They can occur at one--loop level in SM (Standard
Model).  The decay rate is given by~\cite{Pal:1981rm}
\begin{equation}
\Gamma =
\frac{9}{16} \
\frac{\alpha}{\pi} \
\frac{G^2_F}{128 \pi^3} \    
\frac{\left ( \delta m_{ij}^2 \right )^3}{m_i}  
\left | \sum_{\alpha} \ U_{i \alpha}^* \
U_{\alpha j} \
 \left (
\frac{m_\alpha ^2}{m^2_W} \right ) \right |^2
\end{equation}
where $\delta m^2_{ij} = m^2_i - m^2_j$ and $\alpha$ runs over $e, \mu$ and
$\tau$.  When $m_i \gg m_j, m_i \sim O (eV)$ and for maximal mixing
$(4U^{*2}_{i \alpha} U^2_{\alpha j}) \sim O(1)$ and $(\alpha \cong \tau)$
one obtains for $\Gamma$
\begin{equation}
\Gamma_{SM} \sim 10^{-45} \ sec^{-1}
\end{equation}
which is far too small to be interesting.  
The decay mode $\nu_i \to \nu_j + \gamma$ comes from an effective
coupling which can be written as:
\begin{equation}
\left (
\frac{e}{m_i + m_j}
\right )
\bar{\psi}_j \sigma_{\mu \nu} (C + D \gamma_5) \psi_i \ F{_{\mu \nu}} 
\end{equation}
Let us define $k_{ij}$ as
\begin{eqnarray}
k_{ij} & = &
\left (
\frac{e}{m_i + m_j}\right ) \sqrt{\mid C \mid ^2  + \mid D \mid^2} \equiv
  k_0 \mu_B  \\ \nonumber
\end{eqnarray}
where $\mu_B = e/2m_e$.  Since the experimental bounds on $\mu_{\nu
  i}$, the magnetic moments of neutrinos, come from reactions such as
$\nu_e e \to e ``\nu''$ which are not sensitive to the final state
neutrinos; the bounds apply to both diagonal as well as transition
magnetic moments and so can be used to limit $k^i_0$ and the
corresponding lifetimes.  The current bounds are~\cite{pdg}:
\begin{eqnarray}
k^e_0 \ < \ 10^{-10} \\ \nonumber
k^\mu_0 \ < \ 7.4 \times 10^{-10} \\ \nonumber
k^\tau_0 \ < \ 5.4 \times 10^{-7}
\end{eqnarray}
For $m_i \gg m_j$, the decay rate for $\nu_i \to \nu_j + \gamma$ is
given by
\begin{equation}
\Gamma = \frac{\alpha}{2 m_e^2} \ m_i^3 \ k_0^2
\end{equation}
This, in turn, gives indirect bounds on radiative decay lifetimes for
$O(eV)$ neutrinos of:
\begin{eqnarray}
\tau_{\nu_{e}} \ > \ 5 \times 10^{18} \ \mbox{sec}  \\ \nonumber
\tau_{\nu_{\mu}} \ > \ 5 \times 10^{16} \ \mbox{sec}  \\ \nonumber
\tau_{\nu_{\tau}} \ > \ 2 \times 10^{11} \ \mbox{sec} 
\end{eqnarray}

We realize that it is the mass eigenstates which have well defined
lifetimes.  Converting these bounds to ones for mass eigenstates would
involve factors of mixing angles squared and with the large angles now
indicated would change the above bounds by factors of 2 to 4.

There is one caveat in deducing these bounds.  Namely, the form
factors C and D are evaluated at $q^2 \sim O (eV^2)$ in the decay
matrix elements whereas in the scattering from which the bounds are
derived, they are evaluated at $q^2 \sim O (MeV^2)$.  Thus, some
extrapolation is necessary.  It can be argued that, barring some
bizarre behaviour, this is justified~\cite{frere}.

\subsection{Invisible Neutrino Decays}
\label{sec:invisible-decays}

\subsubsection{Three-body  Decays}
\label{sec:3-body-decays}

A decay mode with essentially invisible final states which does not
involve any new particles is the three-body neutrino decay mode,
$\nu_i \to 3 \nu$.  In seesaw-type extensions of the standard
electroweak theory these decays are mediated by the
neutral-current~\cite{Schechter:1981cv}, due to the admixture of
isosinglet and isodoublets~\cite{Schechter:1980gr}.  As a result, in
these theories there are nondiagonal couplings of the $Z$ to the
mass-eigenstate neutrinos, even at the tree level
\cite{Schechter:1980gr}.  The neutral current may be expressed in the
following general form
\begin{equation}
P = K^\dagger K  =
\left(\begin{array}{ccccc}
    K_L^\dagger K_L & K_L^\dagger K_H\\
    K_H^\dagger K_L & K_H^\dagger K_H
\end{array}\right)
\label{eq:NC}
\end{equation}
where the matrix $P=P^2=P^\dagger$ is directly determined in terms of
the charged current lepton mixing matrix $K \equiv (K_L, K_H)$.
The different entries in the $P_{LL}$ sector of the $P$ matrix
determine the neutral current couplings of the light neutrinos that
induce their decay. The deviation of $P_{LL}$ from the identity matrix
characterizes the departure from the GIM mechanism in the neutrino
sector~\cite{Schechter:1980gr,Schechter:1981cv}. In seesaw models this
is expected to be tiny, for neutrino masses in the eV range. Although
it can be enhanced in variant seesaw type
models~\cite{NSImodels2,Gonzalez-Garcia:1988rw} where the isosinglet
heavy leptons are at the weak scale instead or possibly even lighter,
this decay is still likely to be negligible.

Another way to induce this decay is through radiative corrections.
Indeed, the $\nu_i \to 3 \nu$ decay, like the radiative mode, can
occur at one--loop level in SM.  With a mass pattern $m_i \gg m_j$ the
decay rate can be written as
\begin{equation}
\Gamma = \frac{\epsilon^2 \ G^2_F \ m_j^5}{192 \pi^3}
\end{equation}
In the SM at one--loop level, with the internal $\tau$ dominating, the
value of $\epsilon^2$ is given by~\cite{Lee:1977ti}
\begin{equation}
\epsilon_{SM}^2 = \frac{3}{16} \left (
\frac{\alpha}{\pi} \right )^2
\left ( \frac{m_\tau}{m_W} \right )^4
\left\{
\ell n
\left (
\frac{m_\tau^2}{m_W^2} \right ) \right \}^2
\left ( U_{\tau j} \ U^*_{\tau i} \right )^2
\end{equation}

With maximal mixing $\epsilon^2_{SM} \approx 3 \times 10^{-12}$.  Even
if $\epsilon$ were as large as 1 with new physics contributions; it
only gives a value for $\Gamma$ of $5 \times 10^{-35} \ sec^{-1}$.
Hence, this decay mode will not yield decay rates large enough to be
of interest.  Although the current experimental bound on $\epsilon$ is
quite poor: $\epsilon < O(100)$, it is still strong enough to make
this mode phenomenologically uninteresting, at least in vacuo.

\subsubsection{Two-body Decays}
\label{sec:two-body-decays}

There is a wide variety of models where neutrinos get masses due to
the spontaneous violation of global lepton number symmetry, leading to
a physical Nambu-Goldstone boson, called majoron. This leads to the
most well-motivated candidate for invisible two-body neutrino
decays~\cite{seesawmajoron,fae}
\begin{equation}
  \label{eq:nnJ}
\nu_{\alpha_{L}} \to \nu_{\beta_{L}} + J   
\end{equation}
All couplings of the majoron vanish with the neutrino masses.  The
structure of the majoron coupling to mass-eigenstate neutrinos
requires a careful diagonalization of the neutrino mass
matrix~\cite{Schechter:1981cv}. When one performs this, typically one
finds that the majoron couling matrix has a strong tendency of being
diagonal. Such GIM-like effect is a generic feature of the simplest
majoron schemes, first noted in Ref.~\cite{Schechter:1981cv}. As a
result, the off-diagonal couplings of the majoron to mass eigenstate
neutrinos relevant for the neutrino decays are strongly
suppressed~\cite{Schechter:1981cv}, so that neutrino decays become
irrelevant~\footnote{ Similarly delicate is the issue of parametrizing
  the majoron couplings.  If one is careful, one can show the full
  equivalence between polar (derivative couplings) and cartesian
  parametrizations~\cite{Dolgov:1996fp}}.

However, majoron couplings are rather model-dependent~\cite{fae}, and
it is possible to contrive models where they are sizeable enough to
lead to lifetimes of phenomenological interest (the first example in
Ref.~\cite{first-fast-decay-paper} is no longer phenomenologically
viable, but it is possible to arrange many variants). 

An alternative way to generate fast invisible two-body neutrino decays
is in models with horizontal symmetries, spontaneously broken at a
scale $\vev\sigma$, instead of lepton number~\cite{Gelmini:1982zz}.
In this case there can be several Goldstone bosons (familons),
characterized by I=0, L=0, J=0.  Even if there is only one familon,
its coupling is typically not subject to the kind of cancellation
characteristic of majoron schemes, so that the new decay mode in
eq.~(\ref{eq:nnJ}) has a decay rate
\begin{equation}
\Gamma =
\frac{g^2_p m^3_\alpha}{16  \pi {\vev\sigma}^2}
\end{equation}
characterized by a dimensionless coupling $g_p$ which is typically
unsuppressed.

In the $SU(2)_L$ symmetry limit one has a similar coupling for the
charged leptons, with corresponding decay modes $\ell_\alpha \to
\ell_\beta + J $.  Thus in this approximation the $\nu_\alpha$
lifetime becomes related to the B.R.  $(\ell_\alpha \to \ell_\beta + J
)$ through
\begin{equation}
\tau_{\nu_{\alpha}} =
\frac{\tau_{\ell \alpha}}{B.R. (\ell_\alpha \to \ell_\beta + J )}
\left (
\frac{m _{\nu_{\alpha}}}{m_{\ell_{\alpha}}} \right )^{-3}
\end{equation}
The current bounds on $\mu$ and $\tau$ branching
ratios~\cite{pdg,Jodidio:1986mz}
\begin{eqnarray}
B.R. (\mu \to e J ) \ < 2 \times 10^{-6} \\ \nonumber
B.R. (\tau \to \mu J ) \ < 7 \times 10^{-6}
\end{eqnarray}
lead to
\begin{eqnarray}
\tau_{\nu_{\mu}} > 10^{24} \  \mbox{sec}  \\ \nonumber
\tau_{\nu_{\tau}} > 10^{20} \  \mbox{sec}.
\end{eqnarray}
These limits also hold for the case of an iso-doublet familon,
$I=1/2$, $L=0$. In addition, one would need to fine tune in order to
avoid mixing with the Standard Model Higgs.

However the $SU(2)_L$ symmetry is broken, so that the above simple
argument is only a very crude approximation.  The strongest direct
bounds on neutrino-neutrino-Goldstone couplings is that which comes
from a study of pion and kaon decays~\cite{Barger:1981vd}, but these
bounds allow couplings strong enough that fast decays are certainly
possible. Similarly the constraint~\cite{Klapdor-Kleingrothaus:1999hk}
which comes from \nbb~\cite{Berezhiani:1992cd}.

From now on we simply assume that fast invisible decays of neutrinos
are possible, and ask ourselves whether such decay modes can be
responsible for any of the observed neutrino anomalies.

We assume a component of $\nu_\alpha,$ i.e., $\nu_2$, to be the only
unstable state, with a rest-frame lifetime $\tau_0$, and we assume
two--flavor mixing, for simplicity:
\begin{equation}
\nu_\alpha = cos \theta \nu_2 \ + sin \theta \nu_1
\end{equation}
with $m_2 > m_1$.  From Eq. (2) with an unstable $\nu_2$, the $\nu_\alpha$
survival probability is
\begin{eqnarray}
P_{\alpha \alpha} &=& sin^4 \theta \ + cos^4 \theta {\rm exp} (-\alpha L/E)
            \\ \nonumber
&+& 2 sin^2 \theta cos^2 \theta {\rm exp} (-\alpha L/2E)
            cos (\delta m^2 L/2E),
\end{eqnarray}
where $\delta m^2 = m^2_2 - m_1^2$ and $\alpha = m_2/ \tau_0$.
Since we are attempting to explain neutrino data without oscillations
there are two appropriate limits of interest.  One is when the $\delta
m^2$ is so large that the cosine term averages to 0.  Then the survival
probability becomes
\begin{equation}
P_{\mu\mu} = sin^4 \theta \ + cos^4 \theta {\rm exp} (-\alpha L/E)
\end{equation}
Let this be called decay scenario A.  The other possibility is when
$\delta m^2$ is so small that the cosine term is 1, leading to a
survival probability of 
\begin{equation}
P_{\mu \mu} = (sin^2 \theta + cos^2 \theta {\rm exp} (-\alpha L/2E))^2
\end{equation}
corresponding to decay scenario B.  

The possibility of solar neutrinos decaying to explain the discrepancy
is a very old suggestion~\cite{Pakvasa:1972gz}.  The most recent
analysis of the current solar neutrino data finds that no good fit can
be found~\cite{Bandyopadhyay:2002qg}; the conclusion is valid for both
the decay scenarios A as well as B.

For atmospheric neutrinos, it was found that for the decay scenario A,
it was not possible to obtain a good fit for all energies.  Turning to
decay scenario B, a reasonable fit was obtained for all the
atmospheric data, with a minimum $\chi^2 = 33.7$ (32 d.o.f.)  for the
choice of parameters
\begin{equation}
\tau_\nu/m_\nu = 63\rm~km/GeV,
\ \cos^2 \theta = 0.30
\end{equation}
The fit is of comparable quality as the one for
oscillations~\cite{atmdecay00}.

The reason for the similarity of the results obtained in the two
models can be understood from the survival probability $P(\nu_\mu \to
\nu_\mu)$ of muon neutrinos as a function of $L/E_\nu$ for the two
models using the best fit parameters is very similar.  In the case of
the neutrino decay model the probability $P(\nu_\mu \to \nu_\mu)$
monotonically decreases from unity to an asymptotic value $\sin^4
\theta \simeq 0.49$.  In the case of oscillations the probability has
a sinusoidal behaviour in $L/E_\nu$.  The two functional forms seem
very different; however, taking into account the resolution in
$L/E_\nu$, the two forms are hardly distinguishable.  In fact, in the
large $L/E_\nu$ region, the oscillations are averaged out and the
survival probability there can be well--approximated with 0.5 (for
maximal mixing).  In the region of small $L/E_\nu$ both probabilities
approach unity.  In the region $L/E_\nu$ around 400~km/GeV, where the
probability for the neutrino oscillation model has the first minimum,
the two curves are most easily distinguishable, at least in principle.
It is entirely possible that the Super-K data and new analysis of this
most recent decay model can eventually rule this out.  K2K and
eventually MINOS can also test this hypothesis~\cite{Barger:1999bg}.

Assuming that the neutrino oscillations provide the most likely
explanation for the bulk of both atmospheric and solar neutrino
observations; is it possible to place limits on the neutrino
lifetimes?  It is obvious that solar neutrino data will provide the
strongest bounds currently possible. It has been argued convincingly
by Beacom and Bell recently that under the most general assuptions the
bound on the lifetime of $\nu_e$ (or the dominant mass eigenstate
components thereof) is $\tau > 10^{-4}$ sec for mass in the eV
range~\cite{Beacom:2002cb}.  The strongest bounds can be obtained in
the future from observation of MeV neutrinos from a Galactic supernova
($\tau \sim10^5 \mbox{sec})$ or high energy neutrinos from AGNs ($\tau
\sim10^3 \mbox{sec})$~\cite{decayUHE}.

\section{$CPT$ and Lorentz Violation}
\label{sec:cpt-lorentz-viol}

\subsection{$CPT$ Violation in Neutrino Oscillations}
\label{sec:cpt-viol-neutr}

Consequences of $CP$, $T$ and $CPT$ violation for neutrino
oscillations have been written down
before~\cite{Schechter:1980gk,Cabibbo:1977nk}. We summarize them
briefly for the $\nu_\alpha\to\nu_\beta$ flavor oscillation
probabilities $P_{\alpha\beta}$ at a distance $L$ from the source.  If
\begin{equation}
P_{\alpha\beta}(L) \neq P_{\bar\alpha\bar\beta}(L) \,,
\qquad \beta \neq \alpha \,,
\end{equation}
then $CP$ is not conserved.  If
\begin{equation}
P_{\alpha\beta}(L) \neq P_{\beta\alpha}(L) \,,
\qquad \beta \neq \alpha \,,
\end{equation}
then $T$-invariance is violated. If
\begin{eqnarray}
P_{\alpha\beta}(L) &\neq& P_{\bar\beta\bar\alpha}(L)\,,
\qquad \beta \neq \alpha \,,
\\
\noalign{\hbox{or}}
P_{\alpha\alpha}(L) &\neq& P_{\bar\alpha\bar\alpha}(L) \,,
\end{eqnarray}
then $CPT$ is violated.  When neutrinos propagate in matter, matter
effects give rise to apparent $CP$ and $CPT$ violation even if the
mass matrix is $CP$ conserving.
The $CPT$ violating terms can be Lorentz-invariance violating (LV) or
Lorentz invariant. The Lorentz-invariance violating, $CPT$ violating
case has been discussed by Colladay and
Kostelecky~\cite{Colladay:1996iz} and by Coleman and
Glashow~\cite{Coleman:1998ti}.

The effective LV $CPT$ violating interaction for neutrinos is of the
form
\begin{equation}
\bar\nu_L^\alpha b_\mu^{\alpha\beta} \gamma_\mu \nu_L^\beta \,,
\label{eq:LV}
\end{equation}
where $\alpha$ and $\beta$ are flavor indices. If rotational
invariance is assumed in the ``preferred'' frame, in which the cosmic
microwave background radiation is isotropic, then the neutrino
energies are eigenvalues of
\begin{equation}
m^2/2p + b_0 \,,
\end{equation}
where $b_0$ is a hermitian matrix, hereafter labeled $b$.
In the two-flavor case the neutrino phases may be chosen such that
$b$ is real, in which case the interaction in Eq.~(\ref{eq:LV}) is
$CPT$ odd. The survival probabilities for flavors $\alpha$ and
$\bar\alpha$ produced at $t=0$ are given by~\cite{Barger:2000iv}
\begin{eqnarray}
P_{\alpha\alpha}(L) &=&
1 - \sin^22\Theta \sin^2(\Delta L/4)\,, \label{eq:foo}\\
\noalign{\hbox{and}}
P_{\bar\alpha\bar\alpha}(L) &=&
1 - \sin^2 2\bar\Theta \sin^2(\bar\Delta L/4) \,,\\ 
\noalign{\hbox{where}}
\Delta\sin2\Theta &=&
\left| (\delta m^2/E) \sin2\theta_m
+ 2\delta b e^{i\eta} \sin2\theta_b \right| \,,
\label{eq:delsin}\\
\Delta\cos2\Theta &=&
(\delta m^2/E) \cos2\theta_m + 2\delta b \cos2\theta_b \,.
\label{eq:delcos}
\end{eqnarray}
$\bar\Delta$ and $\bar\Theta$ are defined by similar equations with
$\delta b\to -\delta b$.  Here $\theta_m$ and $\theta_b$ define the
rotation angles that diagonalize $m^2$ and $b$, respectively, $\delta
m^2 = m_2^2 - m_1^2$ and $\delta b = b_2 - b_1$, where $m_i^2$ and $b_i$
are the respective eigenvalues. We use the convention that
$\cos2\theta_m$ and $\cos2\theta_b$ are positive and that $\delta m^2$
and $\delta b$ can have either sign.  The phase $\eta$ in
Eq.~(\ref{eq:delsin}) is the difference of the phases in the unitary
matrices that diagonalize $\delta m^2$ and $\delta b$; only one of these
two phases can be absorbed by a redefinition of the neutrino states.
Observable $CPT$-violation in the two-flavor case is a consequence of
the interference of the $\delta m^2$ terms (which are $CPT$-even) and
the LV terms in Eq.~(\ref{eq:LV}) (which are $CPT$-odd); if $\delta
m^2 = 0$ or $\delta b = 0$, then there is no observable
$CPT$-violating effect in neutrino oscillations.  If $\delta m^2/E \gg
2\delta b$ then $\Theta \simeq \theta_m$ and $\Delta \simeq \delta
m^2/E$, whereas if $\delta m^2/E \ll 2\delta b$ then $\Theta \simeq
\theta_b$ and $\Delta \simeq 2\delta b$. Hence the effective mixing
angle and oscillation wavelength can vary dramatically with $E$ for
appropriate values of $\delta b$.
We note that a $CPT$-odd resonance for neutrinos ($\sin^22\Theta = 1$)
occurs whenever $\cos2\Theta = 0$ or
\begin{equation}
(\delta m^2/E) \cos2\theta_m + 2\delta b \cos2\theta_b = 0\,;
\end{equation}
similar to the resonance due to matter
effects~\cite{Wolfenstein:1977ue,Barger:2000iv}. The condition for
antineutrinos is the same except $\delta b$ is replaced by $-\delta
b$. The resonance occurs for neutrinos if $\delta m^2$ and $\delta b$
have the opposite sign, and for antineutrinos if they have the same
sign. A resonance can occur even when $\theta_m$ and $\theta_b$ are
both small, and for all values of $\eta$; if $\theta_m = \theta_b$, a
resonance can occur only if $\eta \neq 0$.  If one of $\nu_\alpha$ or
$\nu_\beta$ is $\nu_e$, then matter effects have to be included.

If $\eta=0$, then
\begin{eqnarray}
\Theta &=& \theta \,,
\label{eq:tan}\\
\Delta &=& (\delta m^2/E) + 2\delta b \,.
\label{eq:delta}
\end{eqnarray}
In this case a resonance is not possible. The oscillation
probabilities become
\begin{eqnarray}
P_{\alpha\alpha}(L) &=& 1 - \sin^2 2\theta \sin^2 \left\{ \left( {\delta m^2 
\over 4E} + {\delta b\over 2} \right) L \right\} \,,
\label{eq:P}\\
P_{\bar\alpha\bar\alpha}(L) &=& 1 - \sin^2 2\theta \sin^2 \left\{ \left(  
{\delta m^2 \over 4E} - {\delta b\over 2} \right) L \right\} \,.
\label{eq:Pbar}
\end{eqnarray}
For fixed $E$, the $\delta b$ terms act as a phase shift in the
oscillation argument; for fixed $L$, the $\delta b$ terms act as a
modification of the oscillation wavelength.
An approximate direct limit on $\delta b$ when $\alpha = \mu$ can be
obtained by noting that in atmospheric neutrino data the flux of
downward going $\nu_\mu$ is not depleted, whereas that of upward going
$\nu_\mu$~is depleted~\cite{atm}.  Hence, the oscillation arguments in
Eqs.~(\ref{eq:P}) and (\ref{eq:Pbar}) cannot have fully developed for
downward neutrinos. Taking $|\delta b L/2| < \pi/2$ with $L\sim20$~km
for downward events leads to the upper bound $|\delta b| <
3\times10^{-20}$~GeV; the K2K results can improve this by an order of
magnitude; upward going events could in principle test $|\delta b|$ as
low as $5\times10^{-23}$~GeV.  Since the $CPT$-odd oscillation
argument depends on $L$ and the ordinary oscillation argument on
$L/E$, improved direct limits could be obtained by a dedicated study
of the energy and zenith angle dependence of the atmospheric neutrino
data.

The difference between $P_{\alpha\alpha}$ and $P_{\bar\alpha\bar\alpha}$
\begin{equation}
P_{\alpha\alpha}(L) - P_{\bar\alpha\bar\alpha}(L) =
- 2 \sin^22\theta \sin\left({\delta m^2 L\over2E}\right) \sin(\delta b L) 
\,, \label{eq:deltaP}
\end{equation}
can be used to test for $CPT$-violation. In a neutrino factory, the
ratio of $\bar\nu_\mu \to \bar\nu_\mu$ to $\nu_\mu \to \nu_\mu$ events
will differ from the Standard Model (or any local quantum field theory
model) value if $CPT$ is violated. A 10kT detector, with 10$^{19}$
stored muons, can probe $\delta b$ to a level of $3. 10^{-23}$
GeV~\cite{Bilenky:2001ka}.  Combining KamLAND and solar neutrino data
would probe $\delta b$ to similar levels.  Lorentz invariant $CPT$
violation can arise if e.g. $\delta m_{ij}^2$ and $\theta_{ij}$ are
different for neutrinos and antineutrinos.  Constraints on such
differences are rather weak~\cite{Barger:2000iv}.  Taking advantage of
this, a very intriguing proposal has been made by several
authors~\cite{Murayama:2000hm}.  It was proposed that in the $\nu$
sector, the $\delta m^2$ and mixing are ``conventional'' and nearly
bimaximal; namely $\delta m^2_{23}$ and $\delta m^2_{21}$ lie in the
atmospheric range determined in Sec.~\ref{sec:atmosph-neutr} and in
the LMA-MSW region determined in Secs.~\ref{sec:solar-neutr},
respectively.  In contrast, in the $\bar{\nu}$ sector $\delta m^2_{23}
\sim 0 (eV^2), \delta m^2_{21}$ lies in the atmospheric range and the
mixing is large in the 1-2 sector but small (of order LSND) in 2-3
sector. Then the $\bar{\nu_\mu}-\bar{\nu}_e$ conversion in
LSND~\cite{LSND} is accounted for, and the solar neutrinos are
unaffected as no $\bar{\nu}'s$ are emitted in the sun.  This proposal
can be tested by Mini-Boone seeing LSND effect in $\bar{\nu}_\mu$
beam, but not in the $\nu_\mu$ beam, and by the fact that the $\nu_e$
and $\bar{\nu}_e$ oscillations with $\delta m^2_{atm}$ will be very
different (present in former and absent in latter).  For example,
KamLAND~\cite{suzuki} will see no effect in reactor $\bar{\nu}_e's$
even if LMA-MSW is the correct solution for solar $\nu_e's$.  This is
of course at odds with the KamLAND confirmation of the LMA-MSW
solution.  At neutrino factories, (fractional) $CPT$ violating mass
differences and mixing parameters can be probed to a percent
level~\cite{Bilenky:2001ka}. It should be stressed that models which
have different masses for particles and anti-particles only seem
Lorentz invariant (and non-local); however, the neutrino propagators
will also violate Lorentz invariance and so they are actually Lorentz
non-invariant as well~\cite{Greenberg:2002uu}.

After the announcement of KamLAND results, which are in general
agreement with the expectations from LMA-MSW and hence $CPT$
conservation; a modified $CPT$-violating scenario to account for LSND
has been proposed~\cite{Barenboim:2002ah}. The idea is that in the
anti-neutrino sector, instead of the LSND and the atmospheric
splittings, now there are the LSND and the KamLAND splittings.  At the
moment it seems possible to fit the atmospheric data.  The solar mass
difference and the KamLAND mass differences need not be the same,
hence LMA-MSW is not yet established according to the authors.

\subsection{Lorentz Invariance Violation in Neutrino Oscillations}
\label{sec:lorentz-invar-viol}

A general formalism to describe small departures from exact Lorentz
invariance has been developed by Colladay and
Kostelecky~\cite{Colladay:1998fq}.  This modification of Standard
Model is renormalizable and preserves the gauge symmetries.  When
rotational invariance in a preferred frame is imposed, the formalism
developed by Coleman and Glashow~\cite{Coleman:1997xq} can be used.
In this form, the main effect (at high energies) of the violation of
Lorentz invariance is that each particle species $i$ has its own
maximum attainable velocity (MAV), $c_i$, in this frame.  The Lorentz
violating parameters are $c^2_i - c_j^2$.
There are many interesting consequences~\cite{Coleman:1997xq}: evading
of GZK cut-off, possibility of ``forbidden'' processes at high
thresholds e.g. $\gamma \to e^+ + e^-, p \to e^+ + n + \nu, \ \mu \to
\pi + \nu_\mu, \ \mu \to e + \gamma$ etc.  Moreover, even if neutrinos
were massless, the flavor eigenstates could be mixtures of velocity
(MAV) eigenstates and the flavor survival probability (in the two
flavor case) is given by
\begin{equation}P_{\alpha \alpha} = 1- sin^2 2 \theta \sin^2
\left (\frac{\delta c}{2} LE \right )
\end{equation}
where $\delta c= c_1-c_2$.
Identical phenomenology for neutrino oscillations arises in the case
of flavor violating gravity or the violation of equivalence principle
(VEP), with $\delta \gamma \Phi$ replacing $\delta c$.  Here $\Phi$ is
the gravitational potential and $\delta \gamma = \gamma_1-\gamma_2$ is
the difference in the post-Newtonian parameters used to test General
Relativity~\cite{Misner} and which break the equivalence principle.
This mechanism was first proposed by Gasperini and by Halprin and
Leung~\cite{Gasperini:zf,Halprin:1991gs}. It provides a different
realization of the phenomenon of oscillation amongst massless
neutrinos, first proposed in Ref.~\cite{first-NSI-resonance-paper} in
the context of neutrino non-standard interactions, as discussed in
Sec.~\ref{sec:non-stand-inter}.
There are, however, some important theoretical differences between the
two proposals. There does not seem to be a consistent theoretical
scheme for VEP, since no theory of gravity obeying the classic General
Relativity tests and also violating the equivalence principle has ever
been found~\cite{Halprin:1995vg}.  In contrast the resonant
oscillation of massless neutrinos due to NSI has a well-defined
theoretical basis, either in terms of effective neutrino
non-orthonomality, or due to the existence of new particles coupled to
neutrinos~\cite{first-NSI-resonance-paper,Nunokawa:1996tg}.
The VEP form of massless neutrino oscillations was very interesting at
one time.  The reason was that a single choice of parameters $\delta
c$ and $\sin^2 2 \theta$ could account for both atmospheric and solar
neutrinos with $\nu_e - \nu_\mu$ mixing~\cite{Pantaleone:ha}.
However, now $\nu_\mu-\nu_e$ can no longer account for atmospheric
neutrinos~\cite{3-nu-sol+atm-fit} and the LE dependence is ruled out
for atmospheric neutrinos~\cite{learned}, except as a sub-leading
effect.
A description of solar neutrinos, even including the recent SNO data,
is still possible~\cite{Raychaudhuri:2001gy}; with the choice of
parameters: $\delta c/2 \sim {10^{-24}}$ and large mixing.  However,
this is ruled out to the extent that KamLAND confirms the LMA-MSW
solution for solar neutrinos; and hence must be a sub-leading effect.
For $\nu_\mu- \nu_x$ mixing, the results of CCFR~\cite{Naples:1998va}
can be used to constrain $\delta c/2 < 10^{-21}$ (for $sin^2 2 \theta
> 10^{-3})$, and future Long Baseline experiments~\cite{minos} will
extend the bounds to $10^{-23}$ for large mixing.
In the general case, when neutrinos are not massless, the energies are
given by
\begin{equation}
E_i= p+ m^2_i/2p + c_i p
\end{equation}
There will be two mixing angles (even for two flavors) and the
survival probability is given by
\begin{equation}
P_{\alpha \alpha}= 1- sin^2 2 \Theta \sin^2 (\Delta L/4)
\end{equation}
where
\begin{eqnarray}
\Delta \sin 2 \Theta & = & \mid(\delta m^2/E) \sin 2 \theta_m + 2 \delta c e^{i \eta}
\ E \sin 2 \theta_c \mid , \\
\Delta \cos 2 \Theta & = & (\delta m^2/E) \cos 2 \theta_m + 2 \delta c 
\ E \cos 2 \theta_c
\end{eqnarray}
One can also write the most general expression including the $CPT$
violating term of Eq. (6) and even extending to three flavors.  But
there is not enough information to constraint the many new parameters.
When data from Long Baseline experiments and eventually neutrino
factories become available, $CPT$ and Lorentz violation in neutrino
oscillations can be probed to new and significant levels.  It would be
especially useful to have detectors capable of distinguishing between
$\nu$ and $\bar{\nu}$ events.

\section{Neutrino Physics with Future Experiments}
\label{sec:impact-kamland-other}

\subsection{Probing Spin Flavor Precession with Borexino}
\label{sec:probing-sfp-with}

\begin{figure}[tbh] 
\centering 
\includegraphics[height=5cm,width=0.46\textwidth]{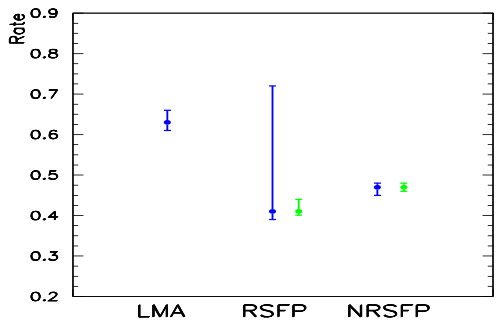}
 \vspace*{-2mm}
 \caption{Predicted $R_{\rm Borexino}$ values for LMA-MSW and
   spin flavor precession solutions, from~\cite{Barranco:2002te}}
\label{fig:boro3.eps}
 \vspace*{-0mm}
 \end{figure}

Irrespective of KamLAND, future data from the upcoming Borexino
experiment will be useful in distinguishing the LMA-MSW solution from
the spin flavor precession solutions described above in
Sec.~\ref{sec:neutr-magn-moments}. Indeed, the Borexino experiment has
the potential to distinguish both the NRSFP solution and the simplest
RSFP solution with no mixing \cite{Barranco:2002te,Akhmedov:2002ti}
from the LMA-MSW solution, as summarized in Fig.~\ref{fig:boro3.eps}.
See Ref.~\cite{Barranco:2002te} for more details.
%%%%%%%%%%%%%%%%%%%%%%%%%%%%%%%%%%%%%%%%%%%%%%%%%%%%%%%%%%%%%%%%%%%%%%
%
 On the other hand a strong confirmation of the LMA-MSW oscillation
 solution by KamLAND~\cite{KamLAND} would imply that
 spin-flavor-precession may at best be present at a sub-leading level,
 leading to a constraint on $\mu B_\perp$.
  
 Note that, if neutrino transition neutrino magnetic moments exist,
 then neutrino conversions within the Sun result will result in
 partial polarization of the initial solar neutrino fluxes.  This
 opens a new opportunity to observe the electron
 antineutrinos~\cite{Pastor:1997pb}. By measuring the slopes of the
 energy dependence of the differential neutrino electron scattering
 cross section one can show how \ne $\to$ \bne conversions may take
 place for low energy solar neutrinos in the Borexino region, while
 being unobservable at the Kamiokande and Super-Kamiokande
 experiments.

\subsection{Probing Spin Flavor Precession with KamLAND}
\label{sec:constraining-mu-b}

\begin{figure*}[tbhp] 
\includegraphics[height=5cm,width=0.96\textwidth]{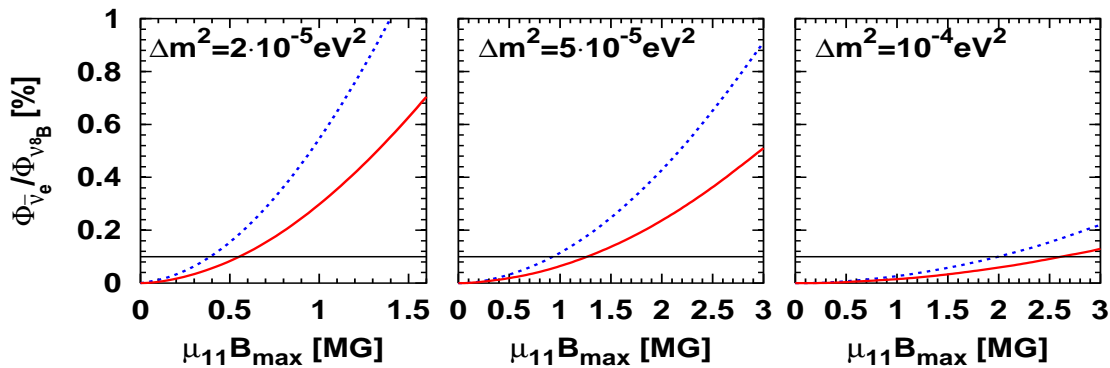}
\caption{KamLAND sensitivity on the Majorana neutrino transition magnetic
  moment for the case of the LMA-MSW solution. See text.}
\label{fig:antikam.eps}
\end{figure*}

Accepting the LMA-MSW solution to the solar neutrino anomaly, as
indicated by first KamLAND results, one can still probe the admixture
of alternative mechanisms of solar neutrino conversion, such as Spin
Flavor Precession. In fact we argue that this will be an interesting
object of study.
With sufficient statistics it should be possible to constrain such
sub-leading admixtures, as discussed in~\cite{Barranco:2002te}.
As an illustration, one can place a constraint on $\mu B_\perp$ (here
$B_\perp$ is the maximum transverse solar magnetic field at the
convective zone) by searching for a solar anti-neutrino flux, expected
in the SFP scenarios. This constraint will depend on how good is the
KamLAND determination of the LMA-MSW oscillation parameters, as
illustrated in Fig.~\ref{fig:antikam.eps}.
%
%Fig:antikam.eps
In this Figure we have displayed the electron anti-neutrino flux
predicted at KamLAND ($E>8.3$~MeV) for three different $\Dms$ values
(indicated in the figure) and for $\tan^2\theta_\Sol$ values varying
in the range from 0.3-0.8, as a function of $\mu_{11} B_{\rm max}$,
$\mu_{11}$ being the magnetic moment in units of $10^{-11}$ Bohr
magneton and $B_{\rm max}$ being the maximum magnetic field in the
convective zone. The extremes of the neutrino mixing range correspond
to the solid and dashed lines indicated in the figure, while the
horizontal line corresponds to a KamLAND sensitivity on the
anti-neutrino flux of 0.1~\%, expected with three years
running~\cite{KamLAND}. Clearly the limits on the transition magnetic
moments are sensitive also to the ultimate central $\Dms$ value
indicated by KamLAND, being more stringent for lower $\Dms$ values, as
seen from the left panel. A 10 \% error on $\Dms$ is aimed at by the
collaboration.

\subsection{Constraining Neutrino Magnetic Moments with Borexino}
\label{sec:constr-mu}

Solar neutrino data can also be used to derive stringent bounds on
Majorana neutrino transition magnetic moments
$\mu_{ij}$~\cite{Schechter:1981hw,Kayser:1982br}, irrespective of the
value of the solar magnetic field.
As discussed in Ref.  \cite{Grimus:2002vb} accepting that LMA-MSW
accounts for the solar neutrino data, one can still probe Majorana
neutrino transition magnetic moments: if present they would contribute
to the neutrino--electron scattering cross section and hence affect
the signal observed in Super-Kamiokande.  Assuming that LMA-MSW is the
solution of the solar neutrino problem Ref.  \cite{Grimus:2002vb}
constrains neutrino TMs by using the latest global solar neutrino
data.  One finds that \texttt{all} elements of the TM matrix can be
bounded at the same time.  Moreover, Ref.  \cite{Grimus:2002vb} shown
how reactor data play a complementary role to the solar neutrino data.
The resulting combined solar plus reactor bound on TMs is $2 \times
10^{-10} \mu_B$ at the 90\% C.L.

\begin{figure}[hbht] \centering
\includegraphics[width=0.5\textwidth,height=6cm]{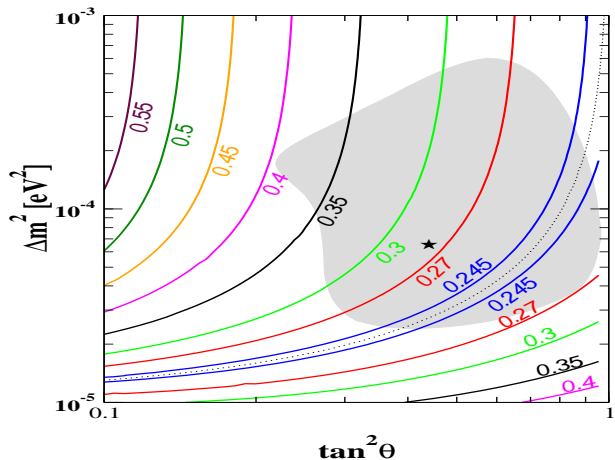}
\caption{Contours of the 90\% \CL\ bound on the magnitude of the Majorana 
  neutrino magnetic moments after 3 years of Borexino data-taking, in
  units of $10^{-10}\mu_B$. The current best fit point is shown by the
  star, and the shaded region is the allowed LMA-MSW region at
  3$\sigma$, from first paper in Ref.~\cite{Grimus:2002vb}. }
\label{fig:borexContour} 
\end{figure}

Contours of the 90\% \CL\ bound on the magnitude of the Majorana
neutrino magnetic moments after 3 years of Borexino data-taking, in
units of $10^{-10}\mu_B$ are displayed in Fig.~\ref{fig:borexContour}.
In this figure the current best fit point is denoted by the star, and
the shaded region is the allowed 3$\sigma$ LMA-MSW region obtained in
the first paper in Ref.~\cite{Grimus:2002vb}, where details of the
analysis can be found.  One sees that, thanks to the lower energy, the
sensitivity of the upcoming Borexino experiment is about an order of
magnitude better than that of current solar neutrino data.
Given the relative delay in the start of the Borexino experiment,
probing neutrino magnetic moments constitutes one of its most
interesting physics goals, as no other current experiment can probe
\TMs with comparable sensitivity. Another interesting item for
Borexino is to test for Non-Standard neutrino interactions. This
possibility has been recently discussed in
Ref.~\cite{Berezhiani:2001rt}.

\subsection{Constraining New Gauge Bosons at Very Low Energies}
\label{sec:constr-new-gauge}

\begin{figure}[tbhp]
\centering
\includegraphics[width=0.45\textwidth,height=5cm]{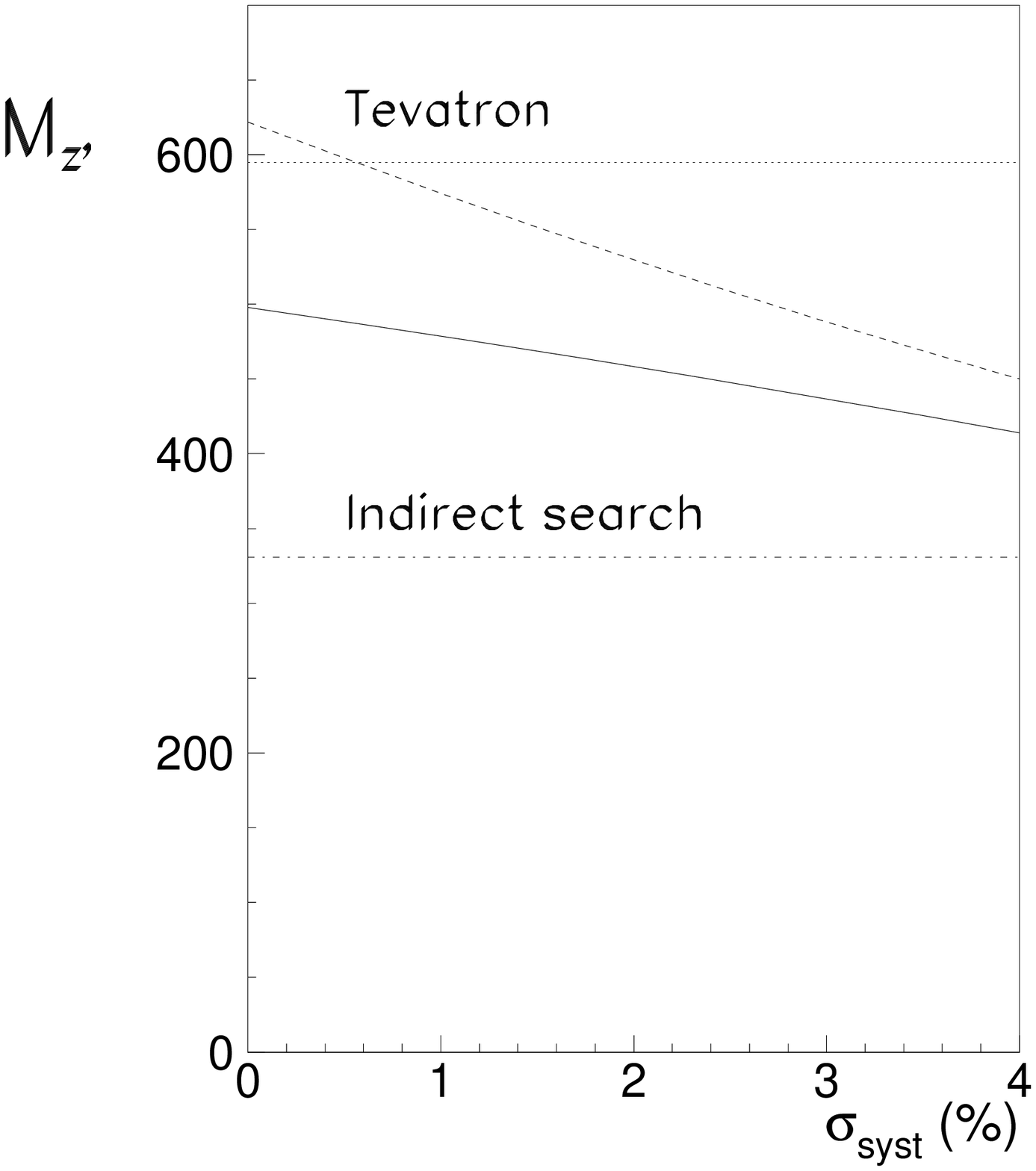}
\includegraphics[width=0.45\textwidth,height=5cm]{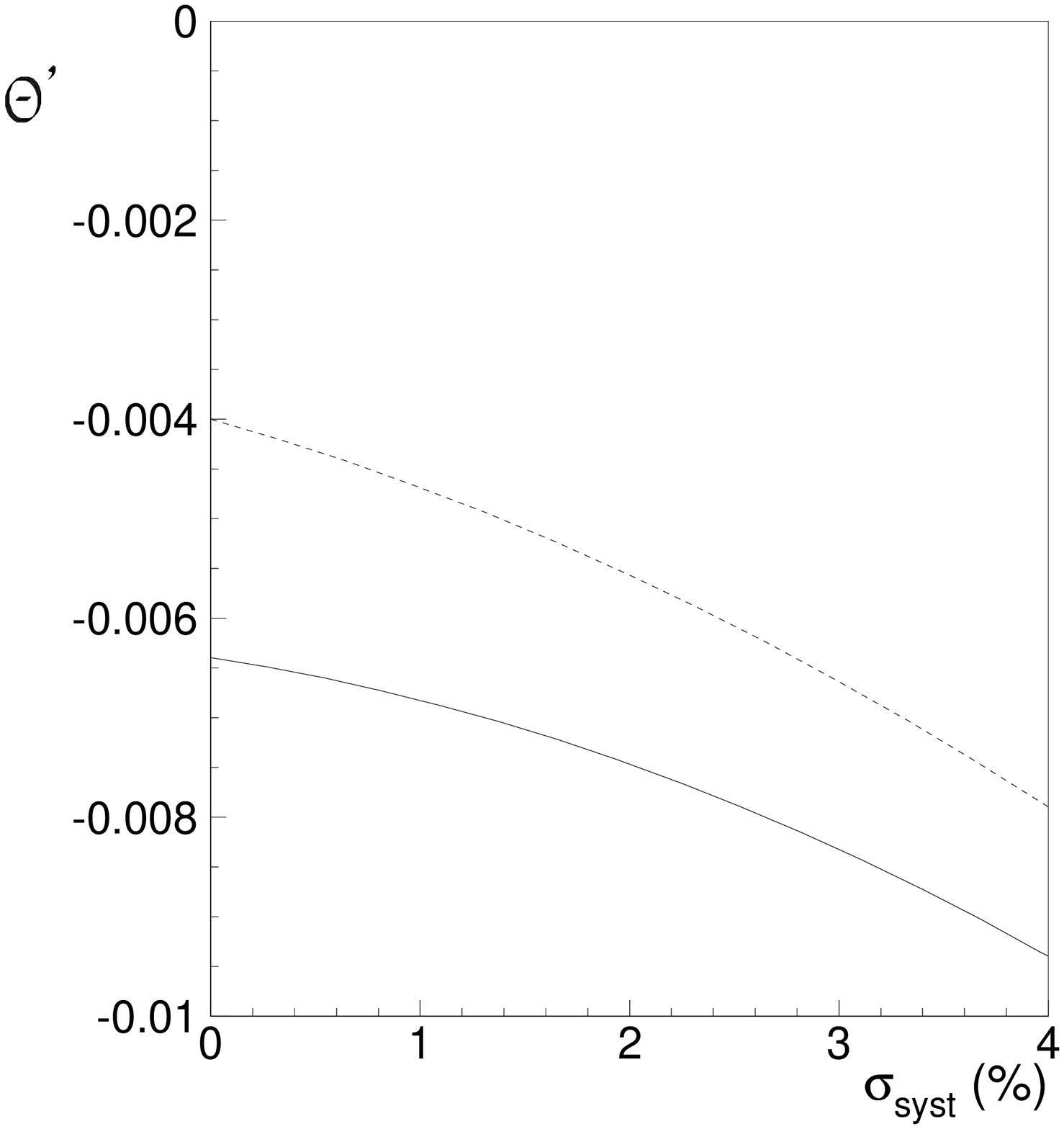}
\caption{Attainable 95 \% C. L. sensitivity 
  to the mass (left panel) and mixing (right) of an extra gauge boson
  in the $\chi$ model, plotted versus the systematic error per bin.
  The results correspond to the cases of a 400 kg (solid line) and
  1-ton detector (dashed line).}
\label{ref:zp}
\end{figure}

Some electroweak models with extended neutral currents, such as those
based on the $E_6$ group~\cite{fae}, lead to an increase of the
$\bar{\nu}-e$ scattering cross section at low energies, typically
below 100 keV~\cite{Miranda:1997vs}. It has been suggested in that the
search for the effects of a heavy Z' gauge boson contribution would be
feasible in an experiment with a high-activity artificial neutrino
source and a large-mass detector.  The neutrino flux is known to
within a percent accuracy and, in contrast to the reactor neutrino
case, one can reach lower neutrino energies. Both features make the
proposed experiment more sensitive to extended gauge models, such as
the $\chi$ model~\cite{fae}. In Fig.~\ref{ref:zp} we briefly summarize
the results obtained in Ref.~\cite{Barabanov:1998bj} for the case for
the proposed LAMA experiment, with a large NaI(Tl) detector located at
the Gran Sasso underground laboratory.
One sees that, for a low enough background, the sensitivity to the
$Z_\chi$ boson mass would reach 600 GeV for one year running of the
experiment. These values are reasonably competitive, and in any case
complementary, to the sensitivity from direct searches at the
Tevatron~\cite{Abe:1997fd}, or through precision electroweak
tests~\cite{precision}.

\subsection{Probing Non-Standard Interactions at Neutrino Factories}
\label{sec:probing-nsi-with}
 
The primary goal of neutrino factories is to probe the lepton mixing
angle $\theta_{13}$ with much better sensitivity than possible at
present and, hopefully, also the possibility of leptonic $CP$
violation~\cite{Apollonio:2002en,Freund:2001ui}.  We have already
discussed both the hierarchical nature of neutrino of mass splittings
indicated by the observed solar and atmospheric neutrino anomalies, as
well as the stringent bound on $\theta_{13}$ that follows from reactor
experiments Chooz and Palo Verde. We also mentioned in
Sec.~\ref{sec:neutr-oscill-param} that the leptonic $CP$ violation
associated to the standard Dirac phase present in the simplest
three-neutrino system (charcterized by a unitary CC mixing matrix)
disappears as two neutrinos become degenerate and/or as $\theta_{13}
\to 0$~\cite{Schechter:1979bn}.  As a result, although the large
mixing indicated by current solar neutrino data certainly helps,
direct leptonic $CP$ violation tests in oscillation experiments will
be a very demanding task for neutrino factories.

\begin{figure}[htb!]
  \begin{center}
    \includegraphics[width=0.65\textwidth]{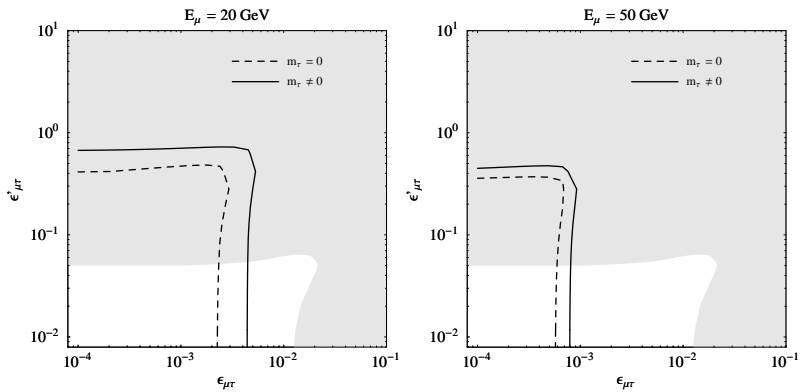}
    \caption{Comparing the sensitivity to neutrino NSI of 
      current atmospheric neutrino experiments (white) with future
      neutrino factory experiments (grey), details in
      Ref.~\cite{Huber:2001zw}}
    \label{fig:sensMTlow}
  \end{center}
\end{figure}

Here we emphasise on the role of neutrino factories in probing
non-standard interactions. It has been shown~\cite{Huber:2001zw} that
NSI in the \nm-\nt channel can be studied with substantially improved
sensitivity in the case of flavor changing NSI, especially at energies
higher than approximately 50 GeV, as illustrated in
Fig.~\ref{fig:sensMTlow}.  For example, a 100 GeV \texttt{Nufact} can
probe FC-NSI interactions at the level of $|\epsilon| < {\rm few}
\times 10^{-4}$ at 99 \% \CL\

\begin{figure}[htb]
   \begin{center}
     \includegraphics[width=0.35\textwidth]{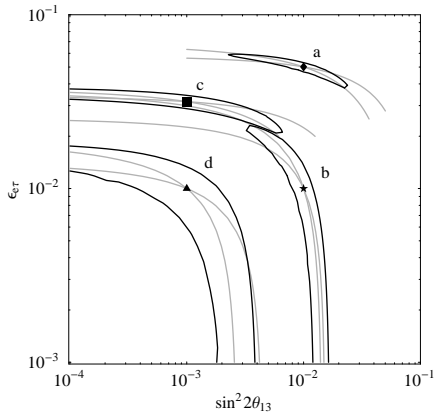}
     \caption{99\% C.L. allowed regions (black lines) in $\sin^2 2\theta_{13}$
       --$\epsilon_{e\tau}$ for different input values, as indicated by
       the points, at a baseline of $3\,000\,\mathrm{km}$.  Lines of
       constant event rates are displayed in grey. Details in
       Ref.~\cite{Huber:2001de}}
     \label{fig:conf}
   \end{center}
 \end{figure}

Note also that in such hybrid solution to the neutrino anomalies, with
FC-NSI explaining the solar data, and oscillations accounting for the
atmospheric data, the two sectors are connected not only by the
neutrino mixing angle $\theta_{13}$, but also by the \ne-\nt flavor
changing NSI parameters. As a result NSI and oscillations may be
confused, as shown in~\cite{Huber:2001de}.  This implies that
information on $\theta_{13}$ can only be obtained if bounds on NSI are
available.  Taking into account the existing bounds on FC
interactions, one finds a drastic loss in \texttt{Nufact}
sensitivities on $\theta_{13}$, of at least two orders of magnitude,
as illustrated in Fig.~\ref{fig:conf}

A near--detector offers the possibility to obtain stringent bounds on
some NSI parameters and therefore constitutes a crucial necessary step
towards the determination of $\theta_{13}$ and subsequent study of
leptonic $CP$ violation.

\section{Discussion and Outlook}
\label{sec:discussion}

These are exciting times for neutrino physics, driven mainly by
experiment.  We have given a brief overview on the status of neutrino
oscillation physics, including the determination of mass splittings
and mixings from current data. Recent results bring substantial
expectation on the potential of future data both of KamLAND and K2K
which provide independent tests of the solar and atmospheric neutrino
anomalies from terrestrial, man-controlled neutrino sources.  The
solar neutrino $\Dms$ will be better determined after 3 years of
KamLAND running, especially if the solar data also improve in the
meantime.  Unfortunately progress in the determination of the solar
angle will be less impressive. In contrast to the Cabibbo angle, the
mixing between the first two generations of leptons has been shown to
be large, though significantly non-maximal. This opens new ways to
probe the solar interior as well as supernovae.

KamLAND has brought a turning point to the possibility of
non-oscillation descriptions of the solar neutrino data, such as those
invoking spin-flavor precession or non-standard neutrino interactions.
Although such descriptions currently provide an excellent fit of the
solar data, they are now globally disfavored by about 3~$\sigma$ and
can only play a sub-leading role in the solar neutrino anomaly.
Analysing in further detail the resulting constraints is beyond the
scope of this short review, which was commissioned well-before these
results were available.

Accepting the LMA-MSW solution, we have nevertheless illustrated how
current and future neutrino data can place important restrictions on
non-standard neutrino properties, such as magnetic moments.
We have also discussed the potential in constraining non-standard
neutrino properties of future data from experiments such as KamLAND,
Borexino and the upcoming neutrino factories.

We have also considered how solar and atmospheric neutrino data can be
used to place constraints on neutrino instability, as well as $CPT$ and
Lorentz Violation. Let us stress that interest still persists in the
investigation of non-standard neutrino properties, to the extent that,
at least some, are well-motivated by theory.

Last, but not least, non-oscillation phenomena such as neutrinoless
double beta decay would, if discovered, probe the absolute scale of
neutrino mass and also reveal their Majorana nature.
With the era of neutrino properties entering a new age, we can only
hope that the underlying mechanism for generating neutrino mass
will start revealing its nature, a formidable task indeed.

\section*{Acknowledgements}
\label{Ackowledgements}

This work was supported in part by Spanish grant BFM2002-00345, by the
European Commission RTN grant HPRN-CT-2000-00148, by the ESF
\emph{Neutrino Astrophysics Network}, and by the U.S.  D.O.E. under
grant DE-FG03-91ER40833.  We thank all our collaborators, for the
physics and for the fun, especially those directly involved on the
analysis of recent neutrino data.  Special thanks also to Michele
Maltoni and Timur Rashba, for technical help, to Sergio Pastor and
Georg Raffelt for discussions on Sec.~\ref{sec:oscill-cosm}, and to
Mariam Tortola for proof-reading.  One of us (S.P.)  would like to
thank the Theory group at KEK for their hospitality while this work
was carried out.

%%%%%%%%%%%%%%%%%%%%%%%%%%%%%%%%%%%%%%%%%%%%%%%%%%%%%%%%%%%%%%%%%%%%%%

\renewcommand{\baselinestretch}{1.00}

\end{document}